\documentclass[pdftex]{ametsocV6.1}
\usepackage{gensymb}
\usepackage{natbib}

\newif\ifdetail
\global\detailfalse

\newcommand{\fig}  [1]{Fig.~\ref{#1}}   
\newcommand{\beq}  { \begin{eqnarray} }
\newcommand{\eeq}  { \end{eqnarray}}
\newcommand{\beeq}  { \begin{eqnarray*} }
\newcommand{\eeeq}  { \end{eqnarray*}}
\newcommand{\eq }[1]{Eq.~(\ref{#1})}  

\newcommand{\E}{\mathcal{E}}

\newcommand{\p}{{\partial}}

\renewcommand{\v}[1]{{\mbox{\boldmath$ #1 $}}} 

\title{Evolution of internal gravity waves in meso-scale eddies}

\authors{Pablo Sebastia Saez\aff{a}\correspondingauthor{Pablo Sebastia Saez, pablo.sebastia.saez@uni-hamburg.de}, Carsten Eden\aff{a}, and  Manita Chouksey\aff{a,b,c}}

\affiliation{\aff{a}{Institut f\"ur Meereskunde, Universit\"at Hamburg, Germany},\\ \aff{b}{Institut f\"ur Umweltphysik, Universit\"at Bremen and MARUM, Germany}, \\ \aff{c}{Leibniz-Institut f\"ur Ostseeforschung Warnem\"unde, Germany} }

\abstract{We investigate the effect of wave-eddy interaction and dissipation of internal gravity waves propagating in a coherent meso-scale eddy simulated using a novel numerical model called the Internal Wave Energy Model based on the six-dimensional radiative transfer equation.
We use an idealized mean flow structure and stratification, motivated by observations of a coherent eddy in the Canary Current System. 
In a spin-down simulation using the Garret-Munk model spectrum as initial conditions, we find that wave energy decreases at the eddy rim. 
Lateral shear leads to wave energy gain due to a developing horizontal anisotropy outisde the eddy and at the rim, while vertical shear leads to wave energy loss which is enhanced at the eddy rim. 
Wave energy loss by wave dissipation due to vertical shear dominates over horizontal shear. Our results show similar behaviour of the internal gravity wave in a cyclonic as well as an anticyclonic eddy.
Wave dissipation by vertical wave refraction occurs predominantly at the eddy rim near the surface, where related vertical diffusivities range from 
$\mathcal{O}(10^{-7})$ to $\mathcal{O}(10^{-5}) \, \rm m^2s^{-1}$.
}
\begin{document}

\maketitle

\statement 
Using a novel model and observations from the Canary Current system of a coherent eddy of 100 km diameter, we explore the interaction between a realistic internal gravity wave field and this eddy. We study wave refraction and energy transfers between the waves and the eddy induced by the eddy’s lateral and vertical shear. Waves lose energy at the eddy rim by vertical shear and gain outside of the eddy rim by horizontal shear. We find large vertical wave refraction by vertical shear at the eddy rim, where waves break and mix density, which can have wide ranging effects on the ocean's circulation. These results are important to understand the ocean’s mixing and energy cycle, and to develop eddy and wave parameterisations.

\section{Introduction}

Internal gravity waves (IGW) are ubiquitous in the ocean and their spectra are observed 
to be quite continuous in wavenumber and frequency space, as described by 
the Garrett-Munk model spectrum \citep{Garrett1975S,Garrett_Munk_1979}. 
IGWs are generated by a variety of mechanisms, primarily by winds and tides \citep{wunschferrari2003}, and also by meso-scale eddies \cite[e.g.][]{ferrari_2011,chouksey_2018,chouksey_2022} in the ocean. Although IGWs evolve over fast time scales, their propagation and associated energy transfers 
can occur over large spatial scales, ranging from a few meters to hundreds of kilometers. 
Interactions between IGWs and meso-scale eddies, which evolve   at much slower time scales, are important for the energy transfers in the ocean, however, these interactions make it challenging to separate, diagnose and quantify the associated oceanic processes.
This study focuses on the evolution of IGWs propagating in a coherent meso-scale eddy using a novel Internal Wave Energy Model (IWEM) based on the six-dimensional radiative transfer equation which allows to separate some of the processes as described below.

During their propagation, IGWs are refracted by their changing environment and can interact with the mean flow, topography, small-scale turbulence, other waves such as Rossby waves, or with IGWs themselves.  For instance, when IGWs propagate in a vertically sheared mean flow, energy can be transferred from the waves to the mean flow and vice versa. Similarly, due to  horizontal shear  waves can exchange energy with the mean flow.
In a critical layer, IGWs get trapped with growing vertical wavenumber (but constant horizontal wavenumber) due to the vertical shear, which eventually leads to wave breaking and thus wave dissipation, sometimes called critical layer absorption.
A similar process is wave capture 
\citep{buhler_2005} where horizontal
shear is involved and all three wavenumber components grow without bounds
which may also lead to wave breaking
and dissipation.

During the breaking, the IGW energy is partly converted to small-scale turbulent energy (which is the dissipated wave energy)
and is transferred to smaller scales, where it is converted into both internal energy
by molecular dissipation and potential energy by turbulent density mixing.
This transfer to potential energy by density mixing
is known to be an important driver of the large-scale ocean circulation \citep[e.g.][]{wunschferrari2003, KunzeSmith2004}. 
However, mixing by wave breaking is not only generated by wave dissipation by vertical and horizontal wave refraction due the mean flow.
The energy transport towards waves with smaller vertical wavelengths, which are prone to breaking as a result of resonant triad wave-wave interactions within the IGW field itself, is also thought 
to be an important driver of wave breaking and mixing. 
The importance of vertical and horizontal wave refraction for wave dissipation as compared to other processes, however, remains largely unknown, and is
thus the focus of the present study.

Estimates of IGW energy 
dissipation and turbulent mixing from observational studies \citep[e.g.][]{polzin2004,polzin2009} confirm that the 
complexity of the vertical and horizontal structure of IGW induced mixing 
is not captured by a constant vertical diffusivity, as typically used in ocean models.
Addressing these issues, \citet{mullerbriscoe_2000} and \citet{muller_natarov_2005} proposed a promising concept based on a prognostic spectral energy balance for IGWs, 
called the Internal Wave Action Model (IWAM) with six-dimensions,
with three dimensions in the spatial physical space and three in the wavenumber space (in addition to the time dimension). 
However, no actual application was published from the IWAM model due to analytical, practical, 
and computational limitations owing to the need to parameterize wave generation, energy transfers, and dissipation. 
Consequently, \citet{olbers2013} drastically simplified this six-dimensional problem, 
and by integrating the spectral energy balance over the wavenumber space
an energetically consistent parameterization for diapycnal diffusivity was given \citep{Olbers2019}.

For the spectral energy balance to hold, it is necessary to assume slowly varying mean flow and stratification compared to
wave lengths and periods, i.e. 
the WKBJ approximation.
Recently, \cite{savva_kafiabad_vanneste_2021} derived a kinetic equation 
similar to the spectral energy transfer equation
using the Wigner transform \citep{RYZHIK1996327}
without assumptions of spatial scale separation. However, in contrast to \cite{savva_kafiabad_vanneste_2021},
we use in the present study the spectral energy balance
as in \citet{mullerbriscoe_2000}
(and  also \cite{kafiabad_savva_vanneste_2019}), 
for reasons discussed in the concluding section.
We
resolve the spectral energy balance in its 
full form 
with the novel numerical model called the Internal Wave Energy Model (IWEM),
to understand the interaction between the geostrophically 
balanced mean flow in form of an coherent eddy, with IGWs in the ocean.

In Section \ref{IGW}, we revisit the propagation and refraction of IGWs 
in a mean flow with vertical and meridional shear in the Wentzel-Kramer-Brillouin (WKB) approximation. 
With the novel numerical code IWEM,
we describe the evolution of a continuous wave field by the spectral energy balance
of IGWs which includes transport, refraction, and wave-mean flow interactions within the limits of the WKB approximation.
 We use  the turbulent kinetic equation to infer vertical mixing related to  wave dissipation by vertical and horizontal wave refraction. 
 In Section \ref{simulations}, numerical experiments with IWEM based on observations
 of an coherent meso-scale eddy in the Canary Current System are discussed
  to understand the role of the  horizontal and vertical gradients 
 of the mean flow.
 We close with a summary and discussion of the results.

\section{Energetics of IGWs}\label{IGW}

The ocean can be described as a body of water with characteristic stratification $N$,
where IGWs are superimposed on a mean current $\textbf{U}$ with (usually) large temporal 
and spatial scales in contrast to the shorter periods and wavelengths of the waves. 
Considering a horizontally and vertically sheared horizontal mean flow $\textbf{U}_h(\textbf{x}_h,z)$, 
stratification frequency $N(z)$, and assuming that the temporal and spatial scales 
of $\textbf{U}_h$ and $N$ are large compared to the periods and wavelengths of the waves, 
the WKB approximation can be applied. 
Freely propagating waves with wave ansatz $\exp i(\textbf{k}_{h}\cdot \textbf{x}_{h} + k_zz -\omega_{enc}t)$ 
and wave vector $\textbf{k}=(\textbf{k}_h,k_z)=(k_x,k_y,k_z)$ obey the local dispersion 
relation for the intrinsic frequency $\omega=\omega_{enc}-\textbf{k}_h\cdot \textbf{U}_h$ of IGWs 
\begin{equation}  \label{eq2.7}
    \left(\omega_{enc} - \textbf{U}_h\cdot \textbf{k}_h\right)^{2}=
    \omega(\textbf{k}_{h},k_z,\textbf{x}_{h},z,t) ^{2}=
    \frac{N^{2}k_{h}^{2}+f^{2}k_z^{2}}{k_{h}^{2}+k_z^{2}} 
\end{equation} 
where $\omega_{enc}$ is the frequency of encounter (sometimes also called absolute frequency). 
The intrinsic frequency $\omega$ follows from the wave properties of IGWs propagating 
in a motionless background and is observed by moving along with the mean current. 
The frequency of encounter $\omega_{enc}$ is noticed by a stationary observer 
and is shifted by $\textbf{k}_h\cdot\textbf{U}_h$, the Doppler shift. 

\subsection{Spectral energy balance}\label{RTE}

The IGW field in the ocean is depicted here by a superposition of wave packets. 
The wave packets have a dominant amplitude, wave vector, and frequency, 
that change according to the familiar ray equations  \citep[see e.g. ][]{olbers_eden_2012}.  
Propagating wave packets can interact with the background stratification $N$ or mean flow $\textbf{U}_h$, 
they can  be influenced by wave-wave interactions and other non-linear processes, 
and there can be forcing or dissipation mechanisms that create new wave packets or destroy the existing ones. 
For instance, IGWs can be excited by wind-driven vertical velocity  fluctuations 
at the base of the surface mixed layer,  
or by horizontal (tidal) currents flowing over bottom topography. 
However, here we exclude forcing, explicit dissipation, 
and non-linear processes and focus on 
progpagation, refraction, and mean-flow interaction in spin-down simulations of a given wave field. 

To describe the energetics of an ensemble of IGWs with energy density 
in space and time $E(\textbf{x}_h,z,t)$  and corresponding  wave action  density
$A=E/\omega$, we define its energy power spectral density $\mathcal{E}$, 
such that $E=\int \mathcal{E} \, \mbox{d} \textbf{k}_h \, \mbox{d} k_z $, and
the corresponding wave action density $\mathcal{A}=\mathcal{E}/\omega$.
From wave action conservation of IGWs follows the spectral balance
\begin{equation}  \label{eq2.3.2}
   \partial_t \mathcal{A} 
   + \nabla_{\textbf{x}_h}\cdot (\dot{\textbf{x}}_{h} \mathcal{A})
   + \partial_z(\dot{z}\mathcal{A})
   +\nabla_{\textbf{k}_h}\cdot (\dot{\textbf{k}}_{h}\mathcal{A}) 
   + \partial_{k_z}(\dot{k_z}\mathcal{A})=\mathcal{S}
\end{equation}
The left hand side of  \eq{eq2.3.2}  describes changes of $\mathcal{A}$  
due to tendency, horizontal and vertical propagation by group velocity 
$(\dot{\textbf{x}}_h,\dot{z})=(\nabla_{\textbf{k}_h}\omega_{enc},\partial_{k_z}\omega_{enc})$, 
and refraction in  horizontal and vertical wave vector components by 
$(\dot{\textbf{k}}_h,\dot{k}_z)=-(\nabla_{\textbf{x}_h}\omega_{enc},\partial_{z}\omega_{enc})$. The term $\mathcal{S}$ on the right hand side 
describes processes violating wave action conservation,
such as wave-wave interaction, forcing, or dissipation.
Reformulating \eq{eq2.3.2} for energy $\mathcal{E}= \omega\mathcal{A}$, 
an additional term arises on the right hand side
\begin{equation} \label{eq2.3.3}
    \partial_t \mathcal{E} + \nabla_{\textbf{x}_h}\cdot (\dot{\textbf{x}}_{h}\mathcal{E})+\partial_z (\dot{z}\mathcal{E})
    +\nabla_{\textbf{k}_h}\cdot (\dot{\textbf{k}}_{h}\mathcal{E}) 
    + \partial_{k_z}(\dot{k}_z\mathcal{E})=\omega\mathcal{S}+ \left( \mathcal{E}/\omega \right)
    \dot{\omega}, 
\end{equation}
where the changes in frequency are defined as:
\begin{equation} \label{eq4}
\dot{\omega}= 
-  \left( \dot{\textbf{x}}_h- \textbf{U}_h \right) \cdot 
(\textbf{k}_h\cdot \nabla_{\textbf{x}_h} \textbf{U}_h) 
-\dot{z}(\textbf{k}_h \cdot \partial_z \textbf{U}_h)
\end{equation} 
with $\dot{\textbf{x}}_h- \textbf{U}_h$ being the instrinsic group velocity.
The term $(\mathcal{E}/\omega) \dot{\omega}$ in \eq{eq2.3.3} is a source of 
mechanical energy gathered along the wave path by interaction of the wave field with the background flow. 
This energy can be transferred in both directions, i.e. from the mean flow to the wave field or the other way around. 
Note that wave-wave interaction and wave forcing as part of $\mathcal{S}$
are neglected in \eq{eq2.3.3}, while dissipation is implicitly present as discussed next.

\subsection{Wave breaking and dissipation}\label{section2.2}

As noted earlier, in this study we exclude any explicit wave dissipation.
On the other hand, dissipation (and thus non-zero 
$\mathcal{S}$ in \eq{eq2.3.3}) is still implicitly present in our 
model by a possible wave energy flux $F_{crit}$  across the large vertical wavenumber threshold $k_z^c>0$ 
\begin{eqnarray}\label{fcrit}
F_{v}= \int \left\{  \begin{array}{ll}
 \left[ \dot{k_z} \mathcal{E} \right]_{k_z = k_z^c}   & \mbox{if} ~\dot{k_z} >0
\\  0  & \mbox{else} 
\end{array} \right\}  \, \mbox{d} \textbf{k}_h
- \int \left\{  \begin{array}{ll}
 \left[ \dot{k_z} \mathcal{E} \right]_{k_z = -k_z^c}   & \mbox{if} ~\dot{k_z} <0
\\  0  & \mbox{else} 
\end{array} \right\} \mbox{d} \textbf{k}_h
\end{eqnarray}
and the flux $F_{h} =  F_{h}^x +F_{h}^y$  across the large horizontal wavenumber threshold $k_h^c>0$
\begin{eqnarray}\label{fcapt}
F_{h}^x &=&\int  \left\{  \begin{array}{ll}
 \left[  \dot{k_x} \mathcal{E} \right]_{k_x = k_h^c}   & \mbox{if} ~\dot{k_x} >0
\\  0  & \mbox{else} 
\end{array} \right\} \mbox{d} k_z\mbox{d} k_y 
-\int  \left\{  \begin{array}{ll}
 \left[  \dot{k_x} \mathcal{E} \right]_{k_x = -k_h^c}  & \mbox{if} ~\dot{k_x} < 0
\\  0  & \mbox{else} 
\end{array} \right\}   \mbox{d} k_z\mbox{d} k_y ~~~~
\end{eqnarray}
and similar for $F_{h}^y$.
We set $\pm k_z^c$  and $\pm k_h^c$ to the maximal
values of the model domain in wavenumber space.
The positive definite energy fluxes $F_{v}$ and $F_{h}$  across the wavenumber thresholds are taken to equal the energy transfer from the waves to small-scale turbulence and thus 
density mixing or further dissipation to internal energy. 
This happens, for example, for the case of wave breaking at critical layers, where waves are refracted with growing vertical wavenumber but constant horizontal wavenumber due to the vertical shear only.  Note, however, that often there is both horizontal and vertical shear, such that it is difficult
to relate the flux $F_v$ only to 
the special case of  critical layer absorption. 
Further, $N(z)$ can also drive a flux
$F_v$ even without any vertical shear.
In the following, we refer to wave dissipation by vertical wave refraction for wave energy
to cross the vertical wavenumber threshold $k_z^c$ generating the flux $F_{v}$, see \eq{fcrit},
and similar for $F_h$ from \eq{fcapt}.  
We will test below if the vertical  wavenumbers (or related Richardson number) of the wave field  at the thresholds are already large (or small) enough to warrant this approach.

To treat the fate of the energy transfer $F_{v}+F_{h}$ to small-scale turbulence, 
we consider the turbulent kinetic energy (TKE) equation.
Under the assumption of stationarity and horizontal homogeneity 
and neglection of the transport terms, the TKE equation simplifies to a local balance  
of production terms and dissipation $\epsilon$, 
i.e. the Osborn-Cox relation (see \cite{Osborn_Cox_1972}) given by
\begin{equation}
    -\overline{\textbf{u}_{h}^{'}w^{'}}\cdot\partial_z  \overline{\textbf{u}}_{h}+\overline{b^{'}w^{'}}-\epsilon  = 0  \label{eq2.4.1}
\end{equation}
The primed terms in \eq{eq2.4.1} stand for small-scale turbulent fluctuations,  
$\overline{\textbf{u}}_{h}$ denotes wave velocity, and $\epsilon$  dissipation of TKE. 
The vertical turbulent buoyancy flux is given by  $\overline{b^{'}w^{'}}$, 
with buoyancy $b=-g\rho/\rho_{0}$ and
density $\rho$. The shear production term 
$\overline{\textbf{u}_{h}^{'}w^{'}}
\cdot\partial_z \overline{\textbf{u}}_{h}$ represents the wave-mean flow energy exchange,
i.e. the flux $F_{v}+F_{h}$.
With the  diffusive parameterisation $\overline{b^{'}w^{'}} = \kappa N^2$, it follows that
\begin{equation}\label{eq2.4.2}
    \kappa=\frac{\gamma}{1+\gamma}\frac{F_{v}+F_{h}}{N^2} 
\end{equation}
where  $\kappa$ is the turbulent diffusivity and $\gamma$ the relative mixing efficiency, (see \cite{Osborn_1980}).
The energy transfer  by wave breaking 
$F_{v}+F_{h}$, 
assumed to be equal to the shear production term $\overline{\textbf{u}_{h}^{'}w^{'}}
\cdot\partial_z \overline{\textbf{u}}_{h}$, can either be dissipated into 
heat or leads to an increase of potential energy by density mixing. 
The fraction of the energy dissipated into heat or converted into potential 
energy by density mixing is represented by the relative mixing efficiency $\gamma$. 
Based on fine scale measurements by \citet{Osborn_Cox_1972}
and e.g. \citet{garanaik_venayagamoorthy_2019},
the relative mixing efficiency is $\gamma\approx 0.2$ for stratified flows such as the ocean. 
Typical values of local turbulent diffusivity in the interior of the ocean  are $\kappa\approx 10^{-5} \, \rm m^{2}s^{-1}$, 
 inferred from typical observational estimates of dissipation 
in the range of $\epsilon \approx  10^{-10}$ to $10^{-9} \, \rm m^{2}s^{-3}$.

\subsection{The Garrett-Munk model spectrum}\label{gm_model}

The Garrett-Munk (GM) model spectrum \citep{Garrett1975S,Garrett_Munk_1979} is a  representation 
of the  energy density spectrum $\mathcal E$ of IGWs derived from observations, 
and is used here as the initial condition for our simulations described below. 
The GM model spectrum describes a horizontally isotropic and vertically symmetric spectrum 
with a spectral power law  of -2 in frequency  and vertical wavenumber. 
There are several different versions of the GM model spectrum, here we use 
the following form
\begin{equation}  \label{N123}
    \mathcal{E}(k_z,\omega,\phi,z) = \frac{E(z)}{2 \pi}A(k_z,\omega)B(\omega)
    ~~,~~
     \mathcal{E}(\textbf{k},z)= 
   \mathcal{E}(k_z,\omega,\phi,z)  \, J^{-1} 
\end{equation}
with wavenumber angle $\phi$ and 
\beq
A(k_z,\omega) = \frac{\tilde{A}(k_z/k_z^{*}(\omega))}{k_z^{*}(\omega)}
~~,~~
\tilde{A}(\zeta)=\frac{n_{a}}{1+\zeta^{2}}
~~,~~
B(\omega)=\frac{n_{b} f}{\omega \sqrt{\omega^{2} - f^{2}}}
\\  
J=\frac{\partial(\textbf{k})}{\partial(k_z,\omega,\phi)}
~~, ~~
k_z^{*}(\omega)=\frac{ \sqrt{N^{2}-\omega^{2}}}{c^{*}}
~~,~~
c^{*}=\int_{-h}^{0} \frac{N(z)}{j^* \pi}  \,dz 
\eeq
where
$J$ denotes the Jacobi determinant for the coordinate transformation $(k_z,\omega,\phi)\to(\textbf{k})$,   
$k_z ^{*}$  the bandwidth of the vertical wavenumber spectrum equivalent to the 
vertical mode number $j^{*}=10$ and $\zeta=k_z/k_z^{*}(\omega)$.
The energy density spectrum $\mathcal{E}$ is factorized with the wavenumber shape function $A$ 
and the frequency  shape function $B$. 
Both shape functions integrate to one by choice of the parameter $n_a$ and $n_b$.
The energy density spectrum is then normalized to the total mechanical energy of the wave field 
in physical space, such that 
\beq 
E(z)=E_{0}N(z)/N(z_0) = \int_{f}^N \int_{-\infty}^\infty  \int_0^{2\pi} \E \, \mbox{d} \omega \mbox{d} k_z \mbox{d} \phi
\eeq
where $E_0$ denotes the total energy content $E_0 $ at $z_0$,
and the factor $N(z)/N(z_0)$, the so-called WKB-scaling. 
We use $E_0 = 3 \times 10^{-3} \, \rm m^2/s^2$ and $z_0=0 \, \rm m$ as classical parameter, 
but note that this is probably an underestimation of the typical wave energy \citep{PEO2017}.
While the original GM spectrum uses analytical expressions for the  parameters $n_a$ and $n_b$,
we choose them such that 
$E(z) = \int_{-\v k_h^c}^{\v k_h^c} \int_{-k_z^c}^{k_z^c} \E \, \mbox{d} \textbf{k}_h \mbox{d} k_z$
at all depths.

\subsection{Instability at wavenumber thresholds}

For the assessment of the stability of the wave field, the shear generated 
by the total wave ensemble given by $\mathcal E$ can be analysed 
and a critical value for the vertical  wavenumber threshold $k_z^c$  calculated. 
Wave instabilities can be assessed by the bulk Richardson number
 $\mbox{Ri}_{\mathrm{Bulk}}=N^2/(\partial_z \v u)^{2}$,  where $\mbox{Ri}_{\mathrm{Bulk}}<1/4$ characterises an 
unstable situation and $\mbox{Ri}_{\mathrm{Bulk}}>1/4$ a stable one. 
Following \cite{Munk19819IW},  $k_z^c$ can be inferred in terms of a critical Richardson number 
sustained by the wave field, so that waves with $|k_z|>k_z^c$ are particularly prone to break,
\begin{equation}
    \mbox{Ri}_{\mathrm{Bulk}}^{-1}=\int_{-k_z^c}^{k_z^c } \mbox{d} k_z \int_f^N \mbox{d} \omega \, Ri^{-1}(k_z,\omega)
     =\int_{-k_z^c }^{k_z^c} dk_z \int_f^N \mbox{d} \omega \, \frac{\omega^{2}+f^{2}}{\omega^{2}N^{2}}k_z^{2}\mathcal{E}(k_z,\omega)
    \label{ri}
\end{equation}
where $Ri(k_z, \omega)$ 
is the spectrum of the Richardson number, for which
 $\partial_z \v u$ is expressed
by the shear spectrum in the second step.

However, the value for $k_z^c$ is not uniquely  given in literature.
Different authors use
corresponding wavelengths from $\lambda=10 \, {\rm m} $ down to $\lambda=1 \, {\rm m} $. 
According to \citet{gargett_1981} and in agreement with \citet{gregg_1987}, 
the IGW spectral shape obtained from measurements in the Northwest Atlantic Ocean 
comprises primarily of IGWs with vertical wavelengths $10\, {\rm m}\leq\lambda< 100 \, {\rm m}$, 
finestructure and decaying internal waves between $1 \, {\rm m}\leq\lambda< 10 \, {\rm m}$, 
and small scale turbulence at centimeter scales. 
\cite{gargett_1981} find that the observed spectrum results in Ri$_{\mathrm{Bulk}}=1$ 
and suggests a cut-off vertical wavelength $\lambda_z^c=10m$. 
\cite{Munk19819IW} predicted Ri$_{\mathrm{Bulk}}^{-1}(k_z^c)=0.5$ with a roll-off at $k_z^c=0.6 \, \rm m^{-1}$ 
(wavelength of about 10 m), showing more agreement with observations by \cite{Sherman_Pinkel_1991}, 
where $\mbox{Ri}^{-1}_{\mathrm{Bulk}} \approx 0.3$ \citep{shermanphd_1989} for spectra close to the model spectrum by \citep{gm76}. 
Moreover, \cite{duda_cox_1989} found Ri$_{\mathrm{Bulk}}$ between one and three from observations in the upper ocean, 
where $\mbox{Ri}_{\mathrm{Bulk}}=1$ indicates high non-linearity of the IGW field 
and the upper bound of $\mbox{Ri}_{\mathrm{Bulk}}=3$ indicated saturated situations with maximum allowable shear.

Here,  $k_z^c = 1 \, \rm m^{-1}$ (or wavelength 6.28 m) is used 
and a zonal and meridional wavenumber threshold of $k_h^c = 0.5 \, \rm m^{-1}$. 
This choice results in a $\mbox{Ri}_{\mathrm{Bulk}}^{-1}=0.9$ under consideration 
of a depth-invariant stratification of $N=5 \times 10 ^{-3} s^{-1}$ and 
the GM model spectrum for $\mathcal E$ from above.
For the case of a depth-dependent exponential stratification $N=N_1(z)$ given by \eq{N1}, 
we find    $\mbox{Ri}_{\mathrm{Bulk}}^{-1}\approx 1.4$ at the surface, 
and  $\mbox{Ri}_{\mathrm{Bulk}}^{-1}\approx 0.2$ in the deep ocean.
These predictions of IGW instability are in partial agreement with \cite{gargett_1981}.
However, $\mbox{Ri}_{\mathrm{Bulk}}^{-1}$ depends on the total wave energy, i.e. on the parameter $E_0$ of the GM model spectrum. 
With larger $E_0$, $\mbox{Ri}_{\mathrm{Bulk}}^{-1}$ will increase and $k_z^c$ will decrease. 
Note that three times larger values than used here for $E_0$  are found by \cite{PEO2017}
from ARGO float observations, i.e. the value for $E_0$ used here is most likely a low bias.

\section{Numerical simulations}\label{simulations}
\begin{figure}[h]
\centering
\includegraphics[width=0.8\textwidth]{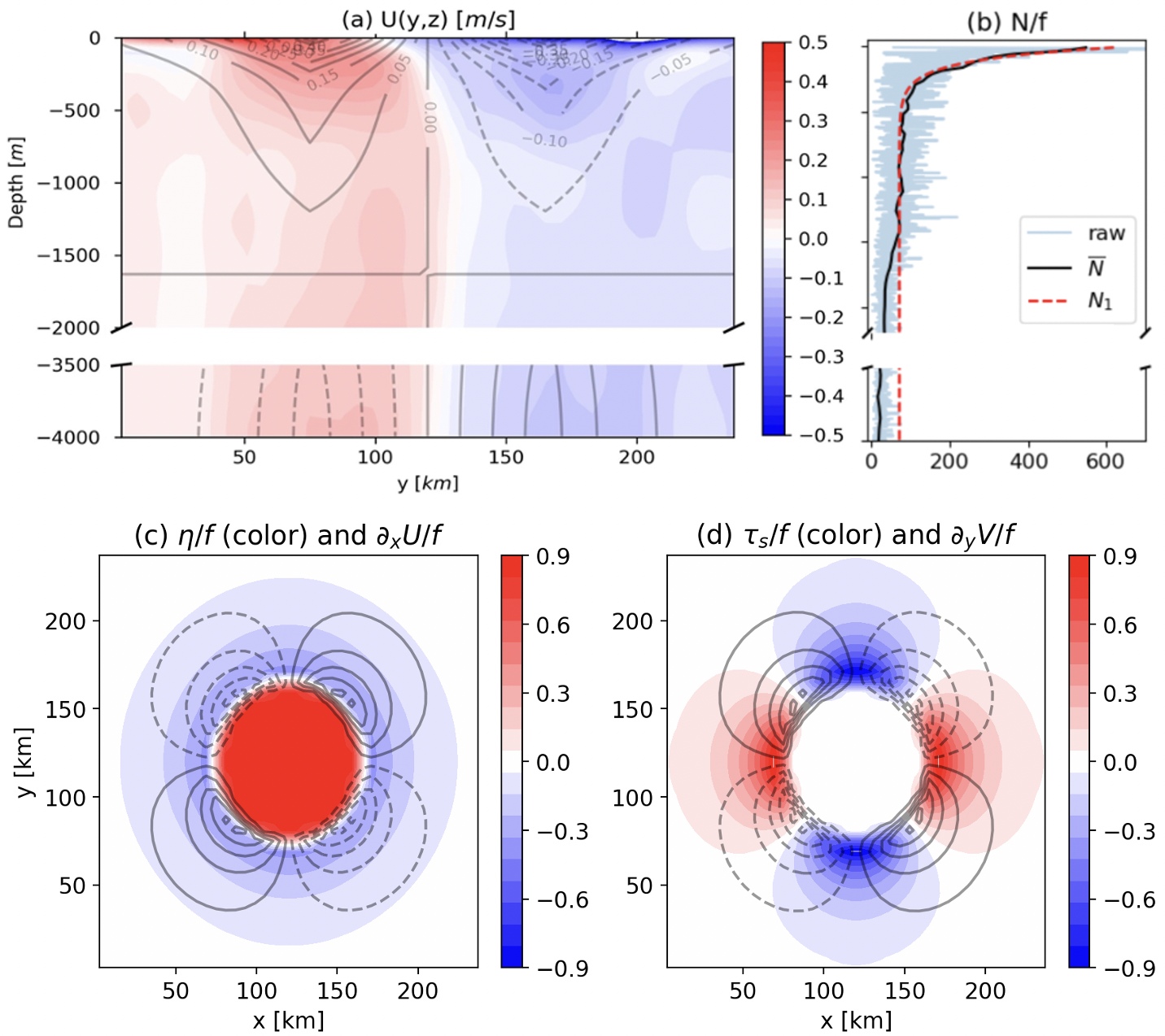}
\caption{(a) Zonal velocity in the coherent eddy in colors calculated from the observed zonal velocity as detailed in Appendix B. 
Grey contours show the zonal velocity of the  idealized eddy  as detailed in Appendix B. (b) Observations of  $N$ (light-blue)
at (14.5\degree N, 24.9\degree W), 
smoothed profile $\overline{N}$ (black) and the respective idealized exponential fit $N_{1}$ (red) 
given by \eq{N1} scaled with the Coriolis frequency $f$.
(c) shows the scaled vorticity $\eta=\partial_{x}V-\partial_{y}U$ in color and zonal shear as contour lines, and (d) shows the shear strain $\tau_{s}=\partial_{x}V+\partial_y{U}$ in color and the meridional shear as contour lines of the idealized eddy close to the surface.
}
     \label{fig:obs_eddy1}
\end{figure}
The numerical simulations in this study are performed with the Internal Wave Energy Model (IWEM), which integrates 
\eq{eq2.3.3} in time. 
A detailed description of the model is provided in Appendix A. 
All numerical experiments are spin-down simulations initialized with the IGW field 
given by the GM model spectrum from Section 2\ref{gm_model}, embedded in a stationary horizontally and vertically varying mean flow 
and a stationary, vertically varying, but horizontally constant stratification.
The integration time is six days, for which we assume the mean flow and stratification to be stationary.
There is no wave forcing, and dissipation only by the energy transports across
the vertical and horizontal wavenumber thresholds as described in Section 2\ref{section2.2}.
The initial wave field  is horizontally homogeneous, isotropic in $\textbf{ k}_h$ and symmetric in $k_z$, 
but  the  total wave energy is
vertically increasing towards the surface since 
the GM-spectrum is proportional to the local stratification $N$ (see Section 2\ref{gm_model}). 

The domain for the simulation has a horizontal extent of $240 \times 240 \, {\rm km}$ and its vertical extent is  $L_z=4\, {\rm km}$ deep.
In physical space, we use $40 \times 40 \times 60$ grid points in the zonal, meridional, and vertical directions 
respectively. The resolution in the horizontal direction is $\Delta x = \Delta y= 6 \, \rm km$. The vertical resolution decreases exponentially from the surface layer with $\Delta z =15 \, \rm m$ to the bottom layer with $\Delta z=74 \, \rm m$.
In wavenumber space we use $60 \times 60 \times 60$ grid points, with horizontal wavenumber $\Delta k_{x,y}=4 \times 10^{-3} \, \rm m^{-1}$ 
at $ k_{x,y}=0 \, \rm m^{-1}$ increasing to 
$\Delta k_{x,y}=18 \times 10^{-3} \, \rm m^{-1}$  at $\pm k_{x,y}^c$, and vertical wavenumber
 $\Delta k_{z}=8 \times 10^{-3} \, \rm m^{-1}$  at $ k_{z}=0 \, \rm m^{-1}$ increasing to 
  $\Delta k_{z}=36 \times 10^{-3} \, \rm m^{-1}$ at $\pm k_{z}^c$.
 The model time step is $2s$.
The total number of grid points is about $20 \times 10^9$, which corresponds roughly
to a global ocean model with horizontal resolution of  1 -- 2 $\rm km$. 
Using even finer grids, the model quickly becomes computationally very expensive such that
the simulations presented here are already at the upper end of available resources to us.

The background flow in 
our model simulations is inspired by measurements of a coherent meso-scale eddy 
  in the tropical northeastern Atlantic Ocean  provided 
by the REEBUS\footnote{\cite{reebus} - Process modelling of physical eddy dynamics} project 
and shown in  \fig{fig:obs_eddy1}  (see also Appendix B),
which we think is representative for a typical 
coherent eddy at mid latitudes.
The eddy measurements were made in the Eastern Boundary Upwelling System at the Mauritanian coast west of Africa in the Canary Current System in December 2019, between  14\degree N to 15\degree N and  24\degree W to 26\degree W and was named $CE\_2019\_14N\_25W$
by REEBUS.
The observed coherent eddy is of cyclonic type with positive zonal  velocity south 
of the eddy center and negative  north of it
(\fig{fig:obs_eddy1}). At the eddy rim, at a distance of around 45 km from the eddy center on either side, 
the  velocities are the largest and reach
$ 0.4 \, \rm ms^{-1}$  
and $ -0.6 \, \rm ms^{-1}$ at the surface.
Between the eddy rim, the horizontal change of the velocity is approximately linear. 
Outside of the eddy, velocity decays exponentially
in the horizontal.
As suggested by  \cite{fischer_2021} this motivates 
 the idealized eddy  structure as detailed in Appendix C, which is
  also shown in \fig{fig:obs_eddy1} (grey contour lines).
  Since the velocity observations reach only to 1000 $\rm m$ depth, 
  we use the first baroclinic vertical mode to extrapolate the observations down to the bottom at 4000 $\rm m$.
We also have a further simulation with an anticyclonic background flow which is simply a mirror of the cyclonic one.
However, if not otherwise noted we always refer to the cyclonic eddy in our discussion
of the results.

The background stratification is computed from in-situ measurements at  the eddy center. 
A mean background stratification $\overline{N}$ is computed by smoothing the observations with a 
moving mean of 150 m  with a minimum of ten data points, shown in 
shown in \fig{fig:obs_eddy1}.
$\overline{N}$ is then fitted with an exponential function $N_{1}(z)$ given by 
 \begin{eqnarray} \label{N1}
    N_{1}(z)&=&  2.5 \times 10^{-3} \, {\rm s^{-1} } + 12.5 \times 10^{-3} \, {\rm s^{-1} }\, \exp{\frac{z+50 \, {\rm m} }{100 \, {\rm m}}  } 
\end{eqnarray}

\begin{figure}[t]
\centering
\includegraphics[width=1.\textwidth]{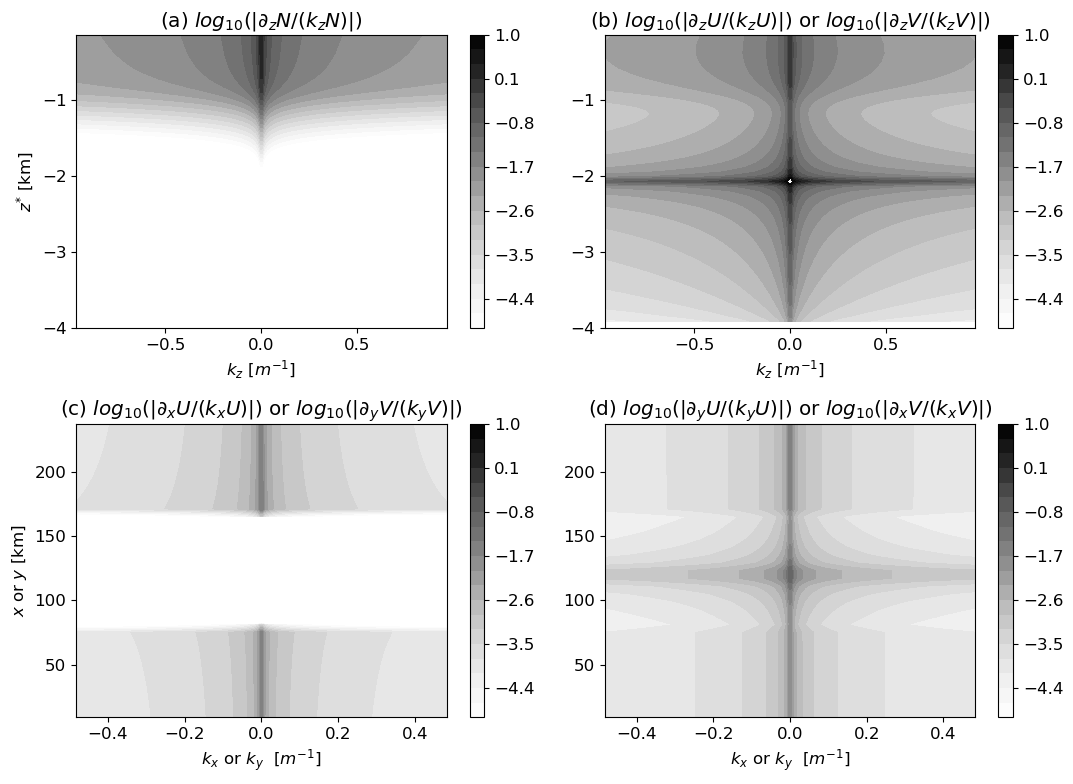}
\caption{The parameters 
$\epsilon=|\partial_z N|/|k_z N|$, 
$|\p_z U|/|k_z U|$, 
$|\p_z V|/|k_z V|$, 
$|\p_x U|/|k_x U|$, 
etc which needs to be small for the WKB approximation to hold. Values for the vertical shear in (a) and (b) are taken at the position of their maximal values 
in $x$ and $y$, and the same holds for (c) and (d) for lateral shear in $z$.
In (b) and (c), the maximum of the two alternatives are shown.} 
  \label{fig:wkb}
\end{figure}
Before we proceed to a discussion of the results of the model simulations, we check
the validity of the WKB approximation, inherent to the model approach here.
For the approximation to hold, it is necessary that the background flow $\v U(\v x,z)$ and stratification $N(z)$ 
are only slowly changing compared to vertical and horizontal wave length,
i.e. $|\p_z N|/ |k_z N|\ll 1$, $|\p_z \v{U}_h |/ |k_z \v{U}_h| \ll 1$, and $|\p_{\v x} \v{U}_h | / |\v{k}_h \v{U}_h| \ll 1 $.
\fig{fig:wkb} shows the maximal corresponding ratios. 
For both $\p_z N$ and $\p_z \v U$ the ratios
are indeed very low, except for very small $k_z$ and the zero crossing of $\v U$ at mid depth.
The ratios for lateral shear stay well below one
everywhere.
We do not regard the large ratios at the smallest $k_z$ as a problem, since here turning 
points are expected which do not show any singular behaviour for the wave energy,
and which are also limited by the finite depth. 
Therefore, the WKB approximation appears to be valid for our model configuration.

\subsection{Waves in a coherent eddy}

\begin{figure}[t]
\centering
\includegraphics[width=1\textwidth]{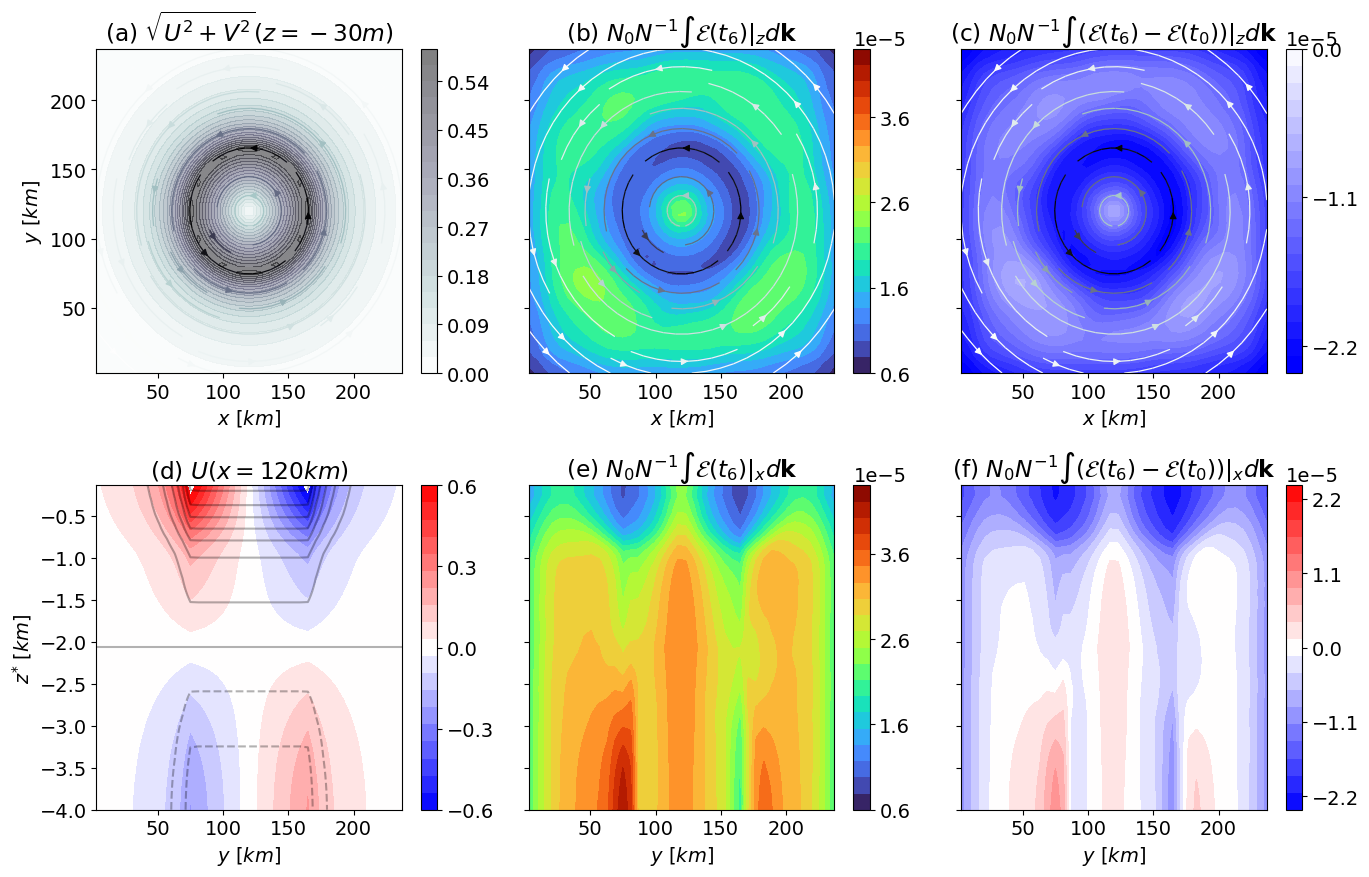}
\caption{(a) Surface speed of cyclonic eddy in $\rm m/s$. (b) Wave energy density $\int \mathcal{E}(t_{6})d\textbf{k} $ in $\rm m^2s^{-2}$ at  $z=-30 \, \rm m$ ($z^*=-266\, \rm m$) and $t_6=6 \, \rm d$. 
(c) Wave energy change over six days in $m^2s^{-2}$. 
The speed of the backgroud flow is shown in  contour lines in panels (a,b,c)
with the same colorscale as in (a).
(d) Meridional cross-section of zonal background velocity $U$ in $\rm m/s$ through the eddy center 
at $x= 120 \, \rm km$
 with meridional velocities in grey contours.
(e)  Wave energy density and wave energy change (f)  over six days. 
Wave energy is WKB-scaled by the factor $N_0N(z)^{-1}$. 
In (d,e,f) we use stretched vertical coordinate $z^*$ with $\mbox{d}z^* = N/N_0 \, \mbox{d} z$,
where $N_0=\int_{-h}^0 (N/h) dz$ with $h=4000 \, \rm m$.
}
 \label{fig:energy}
\end{figure}

We first discuss a simulation  with stratification $N_1(z)$ and the idealized cyclonic coherent eddy. 
The speed of the cyclone is shown in a horizontal plane close to the surface in \fig{fig:energy} (a) 
and a meridional cross-section of zonal velocities of the 
eddy center in \fig{fig:energy} (d). 
We choose to show in \fig{fig:energy} the WKB-scaled wave energy $\mathcal{E}$, i.e. $N_0 \mathcal{E}/N(z)$, 
so that the depth-dependency of $\mathcal{E}$ vanishes at time $t=t_0$. 
We also use the so-called stretched vertical coordinates $z^*(z)$ instead of $z$ 
with $\mbox{d}z^* = N(z)/N_0 \, \mbox{d} z$, where $N_0=\int_{-h}^0 (N(z)/h) dz$ and $h=4000 \, \rm m$, 
to better show the dominant processes close to the surface.

After six days of simulation, we see already a notable change in wave energy in \fig{fig:energy} (b,e). Wave energy has remained relatively stable at the eddy center over the whole water column, while wave energy decreases the most at the eddy rims close to the surface (\fig{fig:energy} b,e). Wave energy changes over time (\fig{fig:energy} c,f) show an overall decrease in wave energy towards the surface, with maxima at the vicinity of the eddy rim and a minimum at the eddy center. 
The waves at the eddy center have gained a small amount of energy 
while IGWs at the eddy rim have lost energy throughout the water column. 
It will be shown below that a similar inhomogeneous distribution of wave energy 
within the coherent eddy can also be seen in the available observations. 
This energy change is the main result of this paper which we aim to understand in the following.

In a simulation with an anticyclonic eddy using the same stratification $N_1(z)$ but simply the negative of the eddy velocities of the cyclone,  the energy changes are almost identical. 
In both spin-down simulations, there is a total wave energy loss of $14.7\%$ relative to the initial energy
integrated over the whole domain.
At the eddy rim there is a relative energy loss of $28.8\%$  integrated
over the water column,
while  the eddy center gains $2.8\%$. 
Note, that the asymmetries in \fig{fig:energy} are due to the finite size effect of simulating the propagation of IGWs in a radial symmetric eddy in Cartesian coordinates. 
The results are symmetric in the eddy core, but the discontinuity in the eddy velocities at the circular eddy rim lead to slight asymmetries
from the eddy rim outwards.

\subsection{Wave-mean flow interaction}
\begin{figure}[t]
 \centering
\includegraphics[width=1\textwidth]{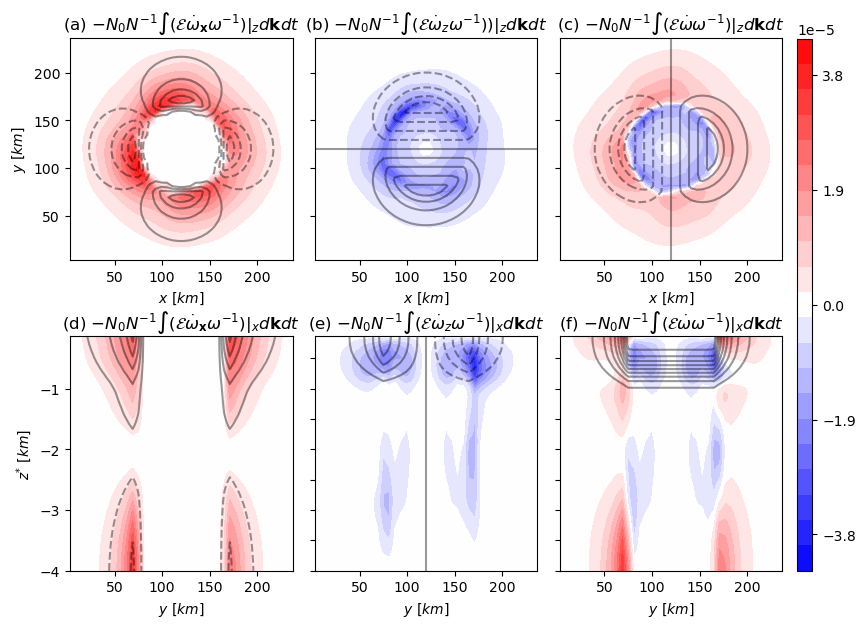}
\caption{Wave energy change by (a) horizontal, (b) vertical and (c) total wave-mean flow interaction  
 (in $[\rm m^{2}s^{-2}]$) after six days of simulation  at  $z=-30 \, \rm m$ ($z^*=-266\, \rm m$).
The grey contours show $\partial_y U+\partial_x V$ in (a),  $\partial_z U$ in (b), and   $\partial_z V$ in (c). 
 (d,e,f) show the same respective components of wave-mean flow interaction 
 and mean flow's shear along the meridional cross-section at $x=120 \, \rm km$.
Positive values in the figure denote a total energy gain.
}
\label{fig:surface_wmf}
\end{figure}

We first consider the wave energy change due
to wave-mean flow interaction.
The  wave-mean flow interaction is given by the term $\mathcal{E}\dot{\omega}\omega^{-1}$ in \eq{eq2.3.3} and depends thus directly on the changes in frequency, see \eq{eq4}. 
We split the discussion into effect of lateral and vertical shear, i.e. 
\beq
\mathcal{E}\dot{\omega}\omega^{-1} = 
\mathcal{E}\dot{\omega}_h\omega^{-1} + \mathcal{E}\dot{\omega}_z\omega^{-1}
\eeq
with $\dot{\omega}_h = - \left( \dot{\textbf{x}}_h- \textbf{U}_h \right) \cdot (\textbf{k}_h\cdot \nabla_{\textbf{x}_h} \textbf{U})$ and  $\dot \omega_z =
-\dot{z}(\textbf{k}_h \cdot \partial_z \textbf{U})$.
All terms are shown in \fig{fig:surface_wmf}.

Consider the effect of lateral shear first.
The horizontal intrinsic group velocity 
can be written as $\dot{\textbf{x}}_h- \textbf{U}_h= v(k,|k_z|) (\cos \phi, \sin \phi) $,
with $\textbf{k}_h = k (\cos\phi, \sin\phi)$
and appropriate function $\nu$.
Then $\dot{\omega}_h $ becomes 
\beq
\dot \omega_h =
 -v k \left(  \cos^2 \phi  \partial_x U +
 \cos \phi  \sin \phi ( \partial_y  U
 + \partial_x V )+
 \sin^2 \phi   \partial_y  V \right) 
\eeq
Since
$\tau_{s}$=$\partial_y U + \partial_x V =0 $, 
 and $\partial_x U = \partial_y V =0 $ within the idealized eddy (\fig{fig:obs_eddy1} c,d),
 it holds that $\dot \omega_h=0$ there,
 and the wave-mean flow interaction by $\mathcal{E}\dot{\omega}_h  \omega^{-1}$ vanishes from the eddy rim towards the center,
 as seen in \fig{fig:surface_wmf} a) and d).
 Outside the eddy rim, the waves 
gain energy due to $\mathcal{E}\dot{\omega}_h  \omega^{-1}$,
which amounts to 6.7\% of the initial wave energy when integrated over  the whole model domain, 
with a maximum close to the eddy rim  near the surface of 48.8\% relative to the initial water column's energy.

Outside the eddy rim, the waves 
gain energy due to $\mathcal{E}\dot{\omega}_h  \omega^{-1}$ because of a developing anisotropic spectrum,
which can be understood as follows:
For an initially isotropic spectrum $\mathcal{E}/\omega = \mathcal{A}(\phi,k,|k_z|)$ with $\partial_\phi \mathcal{A}=0$, we have
\beq
 \int \mathcal{A}\dot \omega_h d k d \phi d k_z = 
- \pi \left(  \partial_x U + \partial_y  V \right) 
\int  \mathcal{A} v k  d k  d k_z = 0 
\eeq
since $ \partial_x U + \partial_y  V =0$ for the idealized eddy,
 $\int_0^{2 \pi}  \cos^2 \phi d \phi=\int_0^{2 \pi}  \sin^2 \phi d\phi = \pi $, and
$\int_0^{2 \pi}  \cos \phi \sin \phi d\phi=0$.
For an isotropic spectrum without $\phi$ dependency, as specified in the initial conditions, the wave-mean flow interaction by lateral shear is thus vanishing
everywhere in the idealized eddy.
Only with anisotropy in $\mathcal{E}$, non-vanishing 
wave-mean flow interaction by lateral shear develops outside the eddy rim, as seen in \fig{fig:surface_wmf} (a,d).

\begin{figure}[t]
 \centering
\includegraphics[width=1\textwidth]{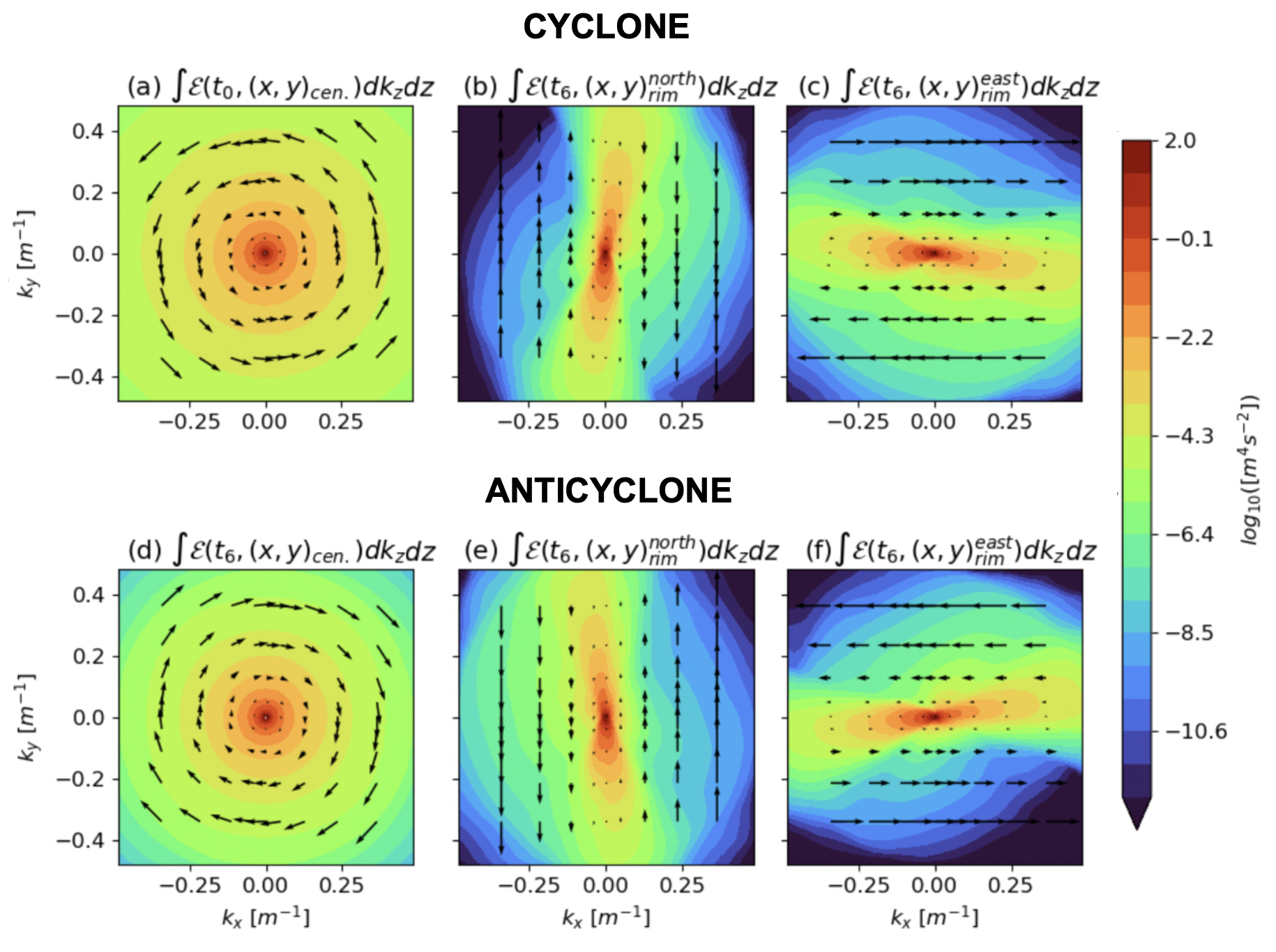}
\caption{Wave energy density integrated over vertical wavenumber $k_z$ and over depth $z$ for the cyclone (top) and anticyclone (bottom): (a) at the eddy center at $t_0$, and after six days of simulation (d) at the eddy center , (b,e) at the northern eddy rim, (c,f) and at the eastern eddy rim . The vector field shows the respective $\dot{\textbf{k}}_h = (k_y,-k_x)\partial_yU $ in (a, d)$, (0,-k_x)\partial_yU $ in (b,e), and $ (-k_y,0)\partial_xV $ in (c,f).}
\label{fig:kxky_anisotropy}
\end{figure}

The horizontal 
anisotropy in $\mathcal{E}$ at the eddy rim 
is generated
in the first place by the effect of vertical refraction, energy transfer to mean flow by vertical shear, and energy dissipation by vertical shear,
which is largest at the eddy rim, and discussed in detail below. 
Using single column simulations (excluding dimensions $x$ and $y$ but not $k_x$ and $k_y$, not shown) with $\v U(z)$ from the eddy rim and the same $N(z)$ as before, i.e. excluding all horizontal processes, we see a
similar horizontal 
anisotropy developing in $\mathcal{E}$.
The effect of vertical refraction, energy transfer to mean flow by vertical shear, and energy dissipation by vertical shear
deforms the initially isotropic 
isolines (circles)  of $\mathcal{E}$  (or $\mathcal{A}$) in the $k_x - k_y $ plane towards ellipses 
with a major axis roughly oriented towards the eddy center with maximal energies inside the ellipse.
This is shown in \fig{fig:kxky_anisotropy}
for the northern and eastern position at the rim.

At the eddy center, however, vertical shear
and thus the effects of  vertical refraction, energy transfer to mean flow, and energy dissipation are vanishing
and no anisotropy is generated by them.
Also the horizontal refraction $\dot{\textbf{k}}_h$ does not generate any anisotropy in $\mathcal{E}$ at the center, since here  $\partial_x U =  \partial_y V=0$ and $\partial_y U =  - \partial_x V$  (\fig{fig:obs_eddy1} c,d), and so 
\begin{equation}
\dot{\textbf{k}}_h = - \nabla_{\textbf{x}_h} \omega_{enc} 
=-\nabla_{\textbf{x}_h} ( \textbf{k}_h\cdot  \textbf{U}_h )
 = (   k_y  ,  - k_x   )  \partial_y U  \sim \textbf{z} \times \textbf{k}_h
\end{equation}  
such that transports by $\dot{\textbf{k}}_h $  are perpendicular to $\textbf{k}_h $ 
with no effect on the isotropic isolines (circles) of $\mathcal{E}$ (see also \fig{fig:kxky_anisotropy} a).
Since the horizontal intrinsic group velocity $\dot{\v x}_h - \v U$
does not depend on wave angle $\phi$ and thus also does not generate any anisotropy from an initially isotropic spectrum, and since for the Doppler shift
$\v U$ in the group velocity the same holds,
$\E$  stays perfectly isotropic 
(\fig{fig:kxky_anisotropy} d).
Only an isotropic  loss of energy due to horizontal wave propagation $\dot{\v x}_h \E$ 
is visible at higher wavenumbers
in \fig{fig:kxky_anisotropy} (d).
This isotropic loss adds to the other effects also at the rim.

An exactly north-south oriented ellipse at the eddy rim is still not effective 
in generating mean-flow interaction by $\dot \omega_h$ since it 
generates an anomaly 
$\mathcal{A}'\sim - \cos 2 \phi $ with
\beq
\int \mathcal{A}' \dot \omega_h  d \phi  \sim 
 - \int v k ( - \cos 2 \phi )
 \cos \phi  \sin \phi ( \partial_y  U
 + \partial_x V ) d \phi = 0
\eeq
The same is true for a general orientation 
of the ellipse towards the eddy center.
However, 
an anomaly with $\sim \sin 2 \phi$ becomes effective at the northern position at the rim.
In fact, if the ellipse is slightly tilted
from north-south direction  direction as seen in \fig{fig:kxky_anisotropy} (b), 
it has a component  $\sim \sin 2 \phi$.
With $\p_x V>0$ but $\p_x V \ll \p_y U$ the transports have indeed 
a tilt in positive $\phi$ direction and an additional anomaly  $\mathcal{A} \sim  -\sin 2 \phi$ develops with
\beq
\int \mathcal{A}' \dot \omega_h  d \phi  \sim 
  \int v k   \sin 2 \phi 
 \cos \phi  \sin \phi ( \partial_y  U
 + \partial_x V ) d \phi = 
 v k  ( \partial_y  U
 + \partial_x V ) \pi /2 >0 
\eeq
which generates gain of wave energy related to $\dot \omega_h$, as seen in  \fig{fig:kxky_anisotropy} (b). 
The same is true for a general orientation 
of the ellipse towards the eddy center,
see e.g. \fig{fig:kxky_anisotropy} (c). 
The slight tilt in positive $\phi$ direction of
the orientation of the anisotropic ellipses in $\mathcal{E}$ thus explains the energy gain outside the eddy rim by the effect of lateral shear.

The small tilt is generated by horizontal refraction 
$\dot{\textbf{ k}}_h$.
North or south of the eddy rim
is $\partial_x U= \partial_y V=0$ and $\partial_y U \gg \partial_x V >0$ (\fig{fig:obs_eddy1} c,d).
The wave refraction is here roughly given  by
\begin{equation}
\dot{\textbf{k}}_h = - \nabla_{\textbf{x}_h} \omega_{enc} 
=-\textbf{k}_h\cdot \nabla_{\textbf{x}_h} \textbf{U}_h
= -(  k_y  \partial_x V,  
k_x  \partial_y U  )   \approx -(0, k_x ) \partial_y U
\end{equation} 
which is also indicated in \fig{fig:kxky_anisotropy} b).
This flow in the $k_x$-$k_y$-plane tilts the 
ellipse slightly to generate the wave energy gain.

In case of the anticyclone,
(\fig{fig:kxky_anisotropy} e and f),
lateral wave-mean flow interaction can be explained analogously, keeping in mind that the horizontal velocities rotate in opposite direction. 
The development of anisotropy as well the orientation of the ellipses towards the eddy center is identical to the cyclone, now however with $\dot{\textbf{k}}_h $ pointing in opposite direction,
which thus generates the slight tilt 
opposite to the one of the cyclonic case.

\fig{fig:surface_wmf} (b,e) shows that  the effect of the vertical shear in 
$\mathcal{E} \dot{\omega}/\omega$ is negative, i.e. the waves lose energy.
This is because of the following reason: The sign of the vertical group  velocity $\dot z$ in 
 $\dot \omega_z =-\dot{z}(\textbf{k}_h \cdot \partial_z \textbf{U}_h)$ depends on the sign of $k_z$.
Since that sign changes when the wave is reflected at the top or bottom boundary  while
all other wave properties are conserved, the effect due to $\dot \omega_z $ is completely 
reversed after reflection, without any net effect after a full cycle of reflection at surface and bottom.
Only dissipation during the wave propagation can break this cycle, and in our model this dissipation is
the energy flux $F_v$ across the vertical wavenumber boundaries.
When the waves propagates towards a critical layer, they always lose energy to the mean flow, thus we
see only wave energy loss by the wave-mean flow interaction related to $\dot{\omega}_z$.
To understand this, consider a wave ray propagating in a vertically sheared mean flow and assume also constant $N$ for simplicity. The wave ray has wave action $A=E/\omega$ starting at time $t_0$ with initially zero mean flow and approaches a critical layer at time $t$. 
Because of wave action conservation $\dot A=0$,
and $\omega(t) = |f| = \omega(t_0) -| U_h k_h | $ $\to$ $E(t)/|f|= E(t_0)/\omega(t_0)$.
Since $|f|/\omega(t_0) = |f|/(|f|+ |k_h U_h |) <1$, wave energy is lost to the mean flow 
during the approach of the ray into the critical layer. 
In the critical layer, the wave breaks and the remainder of the energy is transferred to small-scale turbulence
(or generates secondary waves).
In our model, the remainder of the energy propagates beyond the positive or negative  vertical wavenumber boundary  
which we diagnose as energy flux $F_v$, and which is available for mixing.
Note that only near-inertial waves transport a
significant fraction of energy into the critical layer and to turbulence.

\begin{figure}[h!]
\centering
\includegraphics[width=1\textwidth]{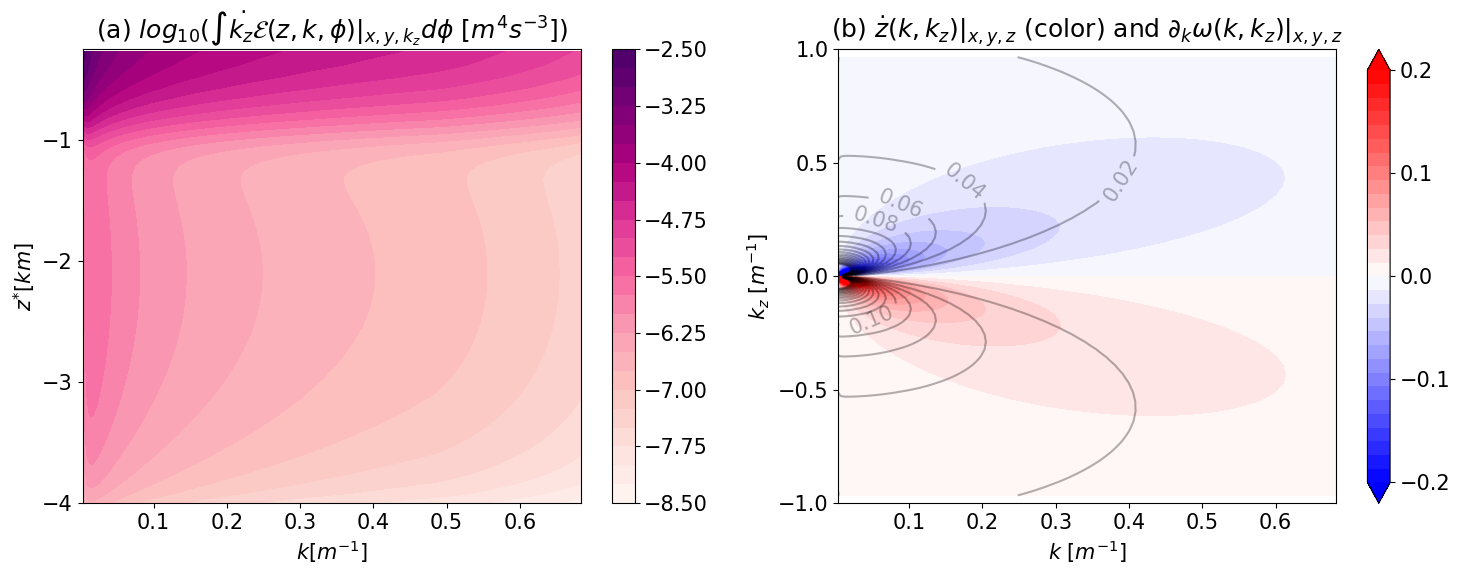}
\caption{(a) Flux $\dot{k_{z}}\mathcal{E}$ 
at $k_z = +k_z^c$ and for $k_{x}>0\, \rm m^{-1}$ integrated over waveangle $\phi$ and  evaluated at the northern eddy rim. (b) Vertical group velocity $\dot z$ (in $[\rm ms^{-1}]$) (color) and intrinsic horizontal group velocity $\partial_{k}\omega(k,k_z)$ (in $[\rm ms^{-1}]$) (contours),
both for $k_{x}>0\, \rm m^{-1}$ and evaluated at the surface at the northern eddy rim.
}   \label{fig:crit}
\end{figure}

We do not have the rather special case of a pure critical layer due to the additional horizontal shear. We, however, observe a ''critical layer like'' wave dissipation by vertical wave refraction, 
such that
the above explanation transfers to our case.
To justify that, \fig{fig:crit} shows 
the outward flux $\dot{k_z}\mathcal{E}$ at the $k_z^{c}$-threshold at the eddy rim where critical layer absorption is strongest (see below).
The outward dissipative flux $\dot{k_z}\mathcal{E}$ is largest at $k \to 0 \, \rm m^{-1}$, i.e. most energy leaves at $\omega \to f$, indicative of a critical layer.
In  \fig{fig:crit} (b) we show the vertical group velocity $\dot z$ (color), which is highest for $k_z \to 0 \, \rm m^{-1}$ and tends to zero for large $k_z$.
The figure also shows 
 that the intrinsic horizontal group velocity $\partial_{k}\omega(k,k_{z})$ (or
 $\dot{\textbf{x}_h}-\textbf{U}_h$)
 also 
tends to zero for large $k_z$ (contours).
We can interpret all these features as  "critical layer like" wave dissipation  behaviour.

Towards the eddy rims, the vertical shear in the mean flow increases,  thus $\dot{\omega}_z$ strengthens, 
with the result that the waves lose more energy to the mean flow there.
 Towards the eddy center $\partial_z \textbf{U}_h \to 0$ and the vertical wave-mean flow interaction vanishes.
The magnitude of the vertical wave-mean flow interaction is increasing towards 
the surface because of the stronger vertical shear of the mean flow. The total relative energy loss by vertical wave-mean flow interaction over
the whole domain amounts to 5\% of the initial wave energy, with  27.3\% relative to the initial  energy at the northern position at the rim.

For the case of the anticyclone, the vertical wave-mean flow interaction is again analogous to the case of the cyclone, keeping in mind, that now the vertical shear in velocities has the opposite sign to the one for the cyclone. Nevertheless, the magnitude of the vertical wave-mean flow interaction increases as well towards the surface due to the stronger vertical shear of the mean flow and leads to a loss of energy, again enhanced at the vicinity of the eddy rim.

The net effect of the wave-mean flow interaction is loss of wave energy from eddy rim to the center
and gain outside of similar magnitude 
(i.e. the eddy would be accelerated inside and
decelerated outside)
such that
both effects appear to cancel each other.
We calculate the integrated net effect of wave-mean flow interaction as 
\beq
WM = 
\frac{ \int \int \E \dot \omega/\omega \, \mbox{d} (\v x,z)  \mbox{d} (\v k,m)dt}
{  \int \int |\E \dot \omega/\omega|  \, \mbox{d} (\v x,z)  \mbox{d} (\v k,m)dt}
\eeq
which is $WM=0.14$ for both cyclone and anticyclone.
There is thus a small net gain of energy by the waves. Wave-mean flow interaction contributes to a total energy gain of $1.7\%$ relative to the initial energy and integrated over the whole cyclone or anticyclone  after the six days of simulation.

\subsection{Wave propagation}
\begin{figure}[t]
 \centering
\includegraphics[width=1\textwidth]{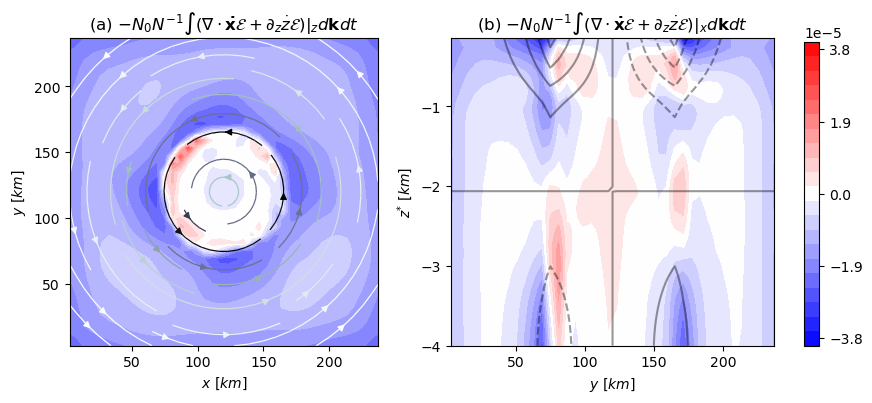}
\caption{ 
Wave energy change due to wave propagation (in $[\rm m^{2}s^{-2}]$) after six days of simulation
(a) at  $z=-30 \, \rm m$ ($z^*=-266\, \rm m$) and (b) over the meridional cross-section at $x=120 \, \rm km$. 
Contour lines show the speed of the eddy in (a) and zonal velocities in (b).
The quantities are WKB-scaled by the factor $N_0N(z)^{-1}$. 
In (b) we use stretched vertical coordinate $z^*$ with $\mbox{d}z^* = N/N_0 \, \mbox{d} z$,
where $N_0=\int_{-h}^0 (N/h) dz$ with $h=4000\, \rm m$.}
\label{fig:surface_prop}
\end{figure}

The total energy change by wave propagation integrated over wavenumber space and the six days of simulation is shown in \fig{fig:surface_prop} for the horizontal plane close to the surface and the meridional cross-section through the eddy. 
The effect of wave propagation is 
as large as the mean-flow interaction and partly counterbalances it,
but is more inhomogeneous in its spatial pattern and hence more difficult to understand.
Horizontal and vertical propagation 
(not shown) contribute with similar magnitudes to the total wave propagation effect in 
\fig{fig:surface_prop}. 
The vertically integrated relative maximal energy loss by wave propagation is $40.6\%$ relative to the initial energy. Wave propagation contributes to a total energy loss of $14.1\%$ relative to the initial energy and integrated over the whole cyclone or anticyclone  after the six days of simulation. 

The intrinsic horizontal group velocity $\dot{ \textbf{x}}_h - \textbf{U}_h$ depends on the wavenumber magnitudes,
but is identical for different wavenumber angles $\phi$.
For the initially isotropic wave field
we thus expect no net effect of the horizontal transport $\dot{ \textbf{x}}_h \mathcal{E}$, except
for a radiation out of the model domain, thus a general decrease of the wave energy.
In principle, this can be seen in $\nabla \cdot \dot{ \textbf{x}} \mathcal{E}$, but  
at the eddy rim, however, there is predominantly wave 
energy gain by $\nabla \cdot \dot{ \textbf{x}} \mathcal{E}$ (not shown)
which is related to the developing anisotropy at the eddy rim.
Outside the eddy and towards the center there is loss of wave energy by the term, with largest magnitudes at the surface and bottom. 
Towards the eddy center, the effect of  
$\nabla \cdot \dot{ \textbf{x} }
\mathcal{E}$ vanishes, since there the wave field stays isotropic.

The effect of  the vertical propagation 
$\partial_z \dot{z} \mathcal{E}$ (not shown) is close to the surface and bottom
 predominantly negative at the rim, and
 positive towards the center, while
 outside the eddy no coherent pattern can be seen.
The net effect by the
wave propagation is a wave energy loss outside the eddy rim related to  wave dissipation by wave refraction (see below),
and small gain or vanishing effect towards the eddy center, see \fig{fig:surface_prop}.
 If it is not a loss, the effect of vertical wave propagation
  tends to be a redistribution of the wave energy over the eddy.
In the vertical propagation effect $\partial_z \dot z \mathcal{E}$, however, 
we see artifacts of the chosen snapshot sampling in time (half daily in our case) which decrease using 
higher snapshot frequencies (which do not show up in other terms), 
but which makes it difficult to understand the role of the term. For the case of the anticyclone, the wave propagation is again analogous to the case of the cyclone.

We use horizontal open boundaries such that waves propagating out of the domain across the horizontal boundaries lead to the largest relative energy loss. 
We assess the role of open boundaries by comparing to an equivalent simulation  with horizontally cyclic boundaries. 
The relative energy loss in time and due to the effect of wave propagation relative to the initial energy integrated over the whole domain is now reduced to approximately $1\%$. However, the effects of wave-mean flow interaction as well as wave dissipation are almost identical for both simulations. Therefore, the waves that propagate out of the domain in the simulations with horizontal open boundaries do not affect our results owith respect to the wave-mean flow interaction and dissipation.

\subsection{Vertical wave refraction and wave dissipation}

 In this section, we discuss the role of  wave dissipation by vertical refraction  in more detail
 and its role for mixing.
The vertical wave refraction is given by 
\begin{equation}\label{kz}
\dot{k_z} = - \partial_z \omega_{enc} =- \textbf{k}_h\cdot \partial_z \textbf{U}_h
-\frac{k_{h}^{2}}{(k_{h}^{2}+k_z^{2})^{2}}\frac{N}{\omega}\partial_z N
\end{equation}
Strong vertical shear $\partial_z \textbf{U}_h$ and large $\partial_z N$  can refract waves 
such that waves eventually  leave the wavenumber domain 
across the vertical  wavenumber threshold  $k_z^c$ 
contributing so to the energy flux $F_v$ ,
which we interpret as wave dissipation by vertical refraction.  $F_v$ contributes to a total of $2.3\%$ of energy loss relative to the initial energy and integrated over the whole cyclone or anticyclone  after the six days of simulation.

The effect of the vertical shear $\partial_z \textbf{U}_h$ dominates the effect of $\partial_z N$ in the
flux $F_v$.
Using single column simulations (excluding dimensions $x$ and $y$ but not $k_x$ and $k_y$, not shown) with $\v U=0$ but the same $\p_z N$,
we see much less wave dissipation by vertical refraction (in this case critical layer absorption) and much smaller flux $F_v$ than including $\p_z \v{U}_h$.
However, the effect of $\partial_z N$ only also drives a small flux $F_v$,
reducing the total energy  by 0.6 \% 
over the six day simulation.

\begin{figure}[t]
 \centering
\includegraphics[width=1\textwidth]{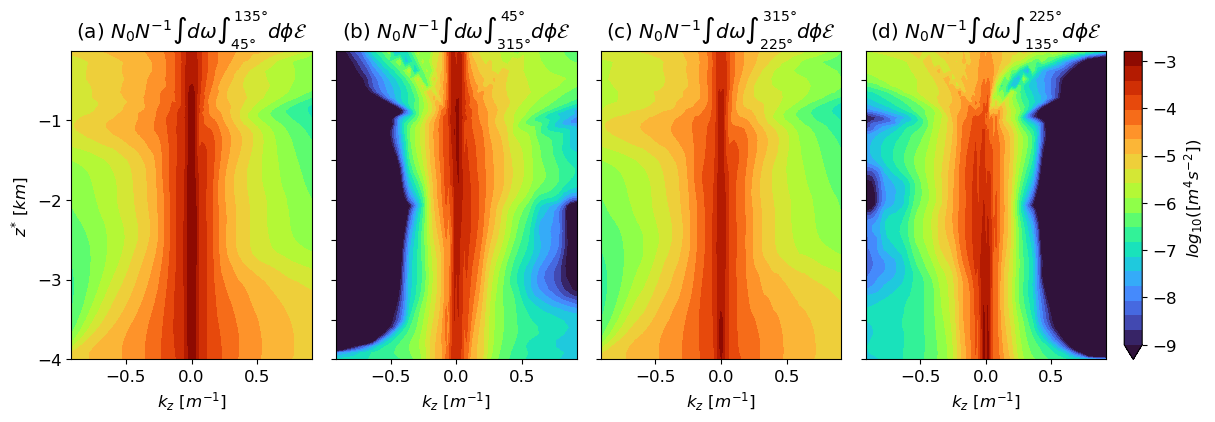}
\caption{
Wave energy density 
at time $t=t_{6}$ in 
$\log_{10}([ \rm m^4/s^2])$  at the northern eddy rim $(x,y)=(120 \, {\rm km},165 \, {\rm km})$ for the wave compartments: (a) north with $45^{\circ} <\phi <135^{\circ}$, (b) east with $315^{\circ} <\phi <45^{\circ}$, (c) south with $225 ^{\circ} <\phi <315 ^{\circ}$ and (d) west with $135 ^{\circ} <\phi <225 ^{\circ}$, where $\phi$ is the angle with respect to the zonal wavenumber ($k_x$) axis. The x-axis in the figure indicates the vertical wavenumber and y-axis the depth.}
\label{fig:compartments}
\end{figure}

Since the dominant effect by  $\partial_z \textbf{U}_h$ in  \eq{kz} depends on the horizontal wavenumber vector $\textbf{k}_h$, it is useful to divide the diagnostics into different wave compartments: north/southward  and west/eastward  propagating waves. That means we perform a coordinate transformation $\mathcal{E}(\textbf{k}_h,k_z,z) \to \mathcal{E}(k_z,\omega,\phi,z)$ and integrate over frequency $\omega$ and the four major cardinal direction, where north is $45^{\circ} <\phi <135^{\circ}$,  east is $315^{\circ} <\phi <45^{\circ}$, south is $225^{\circ} <\phi <315 ^{\circ}$ and west is $135^{\circ} <\phi <225^{\circ}$, where $\phi$ is the angle with respect to the zonal wavenumber ($k_x$) axis. 
We show the compartment-wise integrated energies 
at  the northern eddy rim 
 as function of depth and $k_z$ in \fig{fig:compartments}
 (where $k_z<0$ denotes upward and $k_z>0$  downward propagating waves) after six days of simulation.
At the northern eddy rim is $\partial_{z} U<0$ and $\partial_{z} V \to 0$, i.e. $\dot k_z >0$ 
for eastward propagating waves and 
$\dot k_z<0$ for westward propagating waves
(ignoring the minor effect of $\p_z N$).
Accordingly, \fig{fig:compartments} (b) shows that for the compartment of the eastward propagating waves the upward
propagating branch with $k_z<0$ is emptied, and
wave energy is transported towards the downward propagating branch with $k_z>0$, where the flux 
by $\dot k_z $ across the wavenumber threshold $+k_z^c$ contributes to $F_v$.
That means that east- and downward propagating waves 
dissipate by vertical refraction at the northern rim.
The opposite is true for the westward propagating
waves (\fig{fig:compartments} d), where 
west- and upward propagating waves dissipate by vertical refraction.
The net effect is a drainage of energy in east and westward propagating waves, and no such effect in north and southward propagating waves, generating
the horizontal anisotropy and the ellipse with major axis oriented toward the eddy center seen in \fig{fig:kxky_anisotropy}.
The same holds for the other positions at the rim,
and for the anticylonic eddy accordingly.
The effect of $\p_z N$ on $\dot k_z$ is minor but can also contribute to $F_v$ which can be seen in \fig{fig:compartments} (a) and (c)
where the effect of $\p_z \v{U}_h$ vanishes 
for the north and southward compartments.
Around $1000 \, \rm m$ depth, the maximal 
$\p_z N>$ generates the largest magnitude of $\dot k_z <0$, 
which 
give rise to a small flux $F_v$  
for upward propagating waves.

Horizontal shear can also lead to wave breaking  and dissipation by horizontal refraction . 
Due to the lateral shear,  waves refract according to $\dot{\v k}_h  =-\v k_h \cdot \nabla_{\v x_h} \v{U}_h $ and can eventually leave the wavenumber domain 
across the lateral wavenumber threshold $k_h^c$  and 
contribute to the energy flux $F_h$, 
which we interpret as wave dissipation by horizontal refraction. 
This effect is, however,  two orders of magnitude smaller than  $F_v$
in all our simulations, therefore we do not discuss it in more detail.

\begin{figure}[h]
\centering
\includegraphics[width=1.\textwidth]{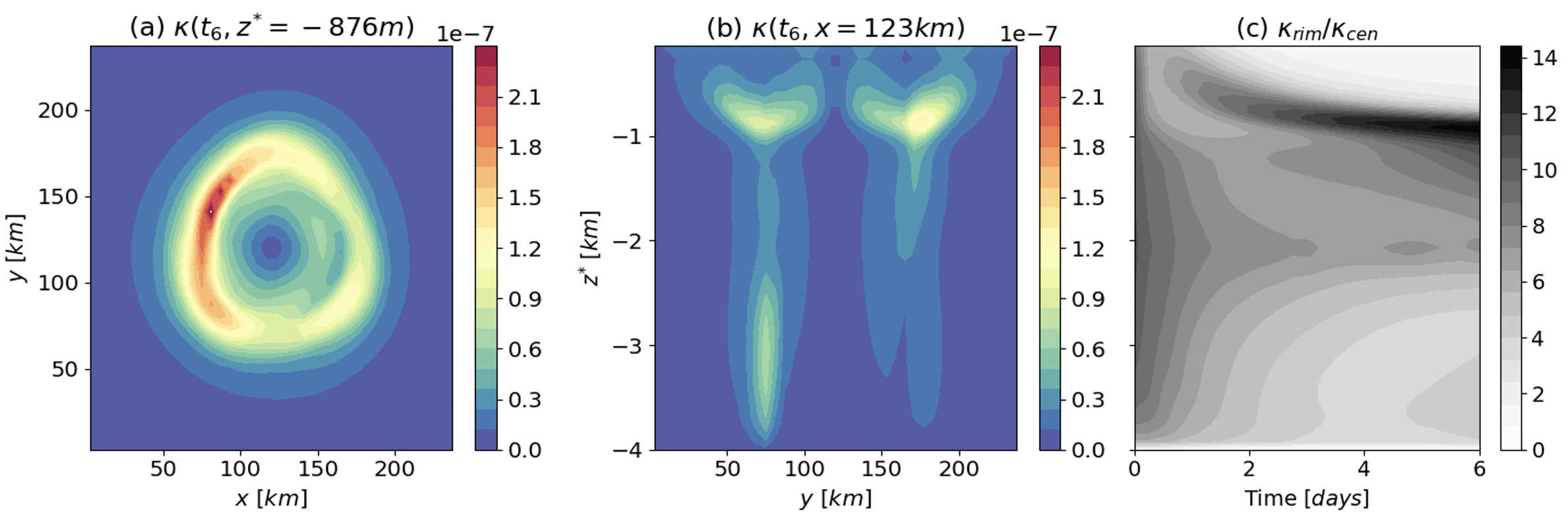}
\caption{Vertical diffusivity $\kappa$ in [$\, \rm m^2s^{-1}$] from wave energy fluxes due to  wave dissipation by vertical refraction $F_v$ (and $F_h$) at  $z=268 \, \rm m$ ($z^*=-876 \, \rm m$) (a), and over the meridional cross-section at $x=120 \, \rm km$ (b) at time $t=t_6$. (c) Ratio of vertical diffusivity between the northern eddy rim and center in time. 
}   \label{fig:diffusivities}
\end{figure}

The vertical diffusivity $\kappa$ is computed from the flux by wave dissipation $F_v$  from \eq{eq2.4.2},
and is shown  in \fig{fig:diffusivities} (a,b)
  after six days simulation. 
  Vertical diffusivities for the case of the cyclonic and anticyclonic eddy are identical.
$\kappa$ is large near the surface  with (decreasing) 
values of $\kappa\approx 10^{-5}$ to $10^{-7} \, \rm m^2s^{-1}$ during the simulation period. 
The diffusivity is maximal at the eddy rim close to the surface 
where most of the wave breaking takes place 
as waves get trapped by the mean shear and dissipate. 
 $\kappa$ is mainly related to $F_v$  with a minor contribution of $F_h$. 
$\kappa$ is maximal at the eddy rim and
 decays towards the eddy center and outside of the eddy as the vertical shear of the mean flow decreases. 

The simulations show much larger diffusivities of the order of $\kappa\approx10^{-5} \, \rm m^2s^{-1}$ towards the surface at the eddy rims during the first two days of simulation (not shown). In \fig{fig:diffusivities} (c) we show the evolution of the fraction between the diffusivities at the northern eddy rim against the center. While at the beginning the diffusivities at the rim are overall larger than at the center, where the shallower maxima in $\kappa$ are predominant (not shown), the lower ones strengthen with time at the rims by more than an order of magnitude.
Eventually, as there is no energy fed to the system, no waves are left to break and no more energy is transported outside of the wavenumber domain, and thus diffusivities vanish. 
\subsection{Comparison to observations}

\begin{figure}[t]
\centering
\includegraphics[width=0.9\textwidth]{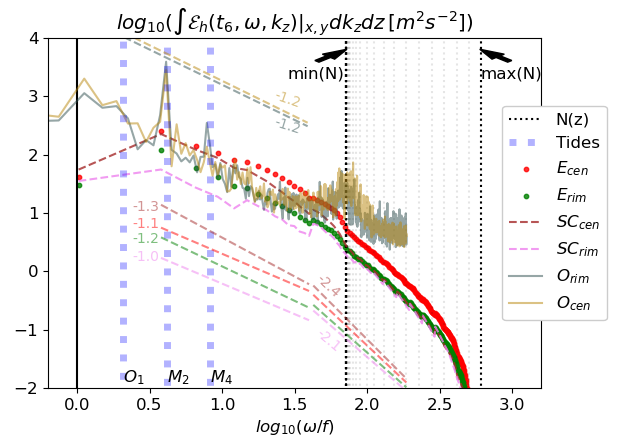}
\caption{Spectra of horizontal kinetic energy densities in frequency space 
at the eddy center and eddy rim in the upper km  from observations ($O_{cen}$ and $O_{rim}$)
and after six days of model simulation ($E_{cen}(t_{end})$ and $E_{rim}(t_{end})$) and of single column simulations ($SC_{cen}$ and $SC_{rim}$). To each spectrum we show the fitted linear regression vertically displaced (dashed lines), where the numbers denote the slopes. 
Dashed blue vertical lines represent tidal frequencies and harmonics ($O_1,M_2...$) 
and dotted grey lines indicate the depth-dependent $N$ where 
waves may encounter turning points. 
Black lines define the frequency domain in which waves live $f\leq \omega \leq max(N)$. Note that the x-axis is scaled with the Coriolis frequency $f$.
 }
           \label{fig:discussion}
\end{figure}

Available observations appear to show the key feature predicted by our model.
A mooring with a downward looking ADCP with $75\, \rm kHz$ 
and a sampling period of $8 \, \rm min$ 
was deployed for 22 days at the core of our eddy within the REEBUS project. 
We divide the measured horizontal velocity time series in two series of six days since then
the first time series corresponds roughly to measurements 
at the eddy core and the second at the eddy rim. 
We compute the horizontal kinetic energy spectral estimate $\mathcal{E}_h(\omega,k_z)$ 
for a moving window of four days for five depth-ranges of $206 \, \rm m$, 
to finally integrate in vertical wavenumber.

\fig{fig:discussion} shows the observed frequency  spectra of horizontal kinetic energy densities 
$\mathcal{E}_h$ at the eddy center and the rim. 
They contain both  elevated energy levels at tidal  frequencies ($O_1$, $M_2$ and harmonics), 
but also all other (wave) frequencies between the local $f$ and $N$ are populated with energy,
akin to the GM model spectrum.
At all such frequencies except for the tidal frequencies, the energy seems indeed to be 
higher for the eddy center than at the rim.
Without more observations from other eddies, it is difficult to say if the difference of the energy level in 
the observations is statistically significant or just by chance.
We made an attempt to proof that the spectral estimate at the eddy center is 
indeed statistically significantly larger compared to the eddy rim by performing 
a $\chi^2$-goodness of fit\footnote{The $\chi^2$-goodness of fit is a statistical method to test two hypotheses: the null and alternative hypothesis. The null hypothesis in this case is: The IGW energy at the eddy center are significantly larger compared to the IGW energy at the rim. The alternative hypothesis is: The IGW energy at the eddy center are not significantly larger compared to the IGW energy at the rim. This statistical test is computed with the Pearson's $\chi^2$ as $\chi^2=\sum_{f} \frac{(O-E)^2}{E}$, where $O$ is the observed variable (in this case the fraction of IGW energy at center against rim for each frequency), $E$ is the expected fraction (in this case, it is $E=1.2$), and $\sum_{f}$ denotes a sum over all frequencies.} of the fraction of the kinetic energy at the center and rim. 
With a simple minimum-finding algorithm we compute an expected fraction 
of $1.2$ that significantly fits ($p>0.05$) our hypothesis. 

However, the difference in the observations clearly has a smaller magnitude than in the model.
The corresponding spectra from the model -- after a coordinate transformation of $\E_h(\textbf{k})$ 
to frequency space $\E_h(\omega,k_z) =\frac{1}{2} \frac{N^2-\omega ^2}{N^2 -f^2}
\frac{\omega^2+f^2}{\omega^2}\mathcal{E}(\omega,k_z)$ --
at $t=0$ and  $t=5 \, \rm d$ for the eddy center and rim  are also shown in \fig{fig:discussion}. Since there is no forcing in the model, and since we initialized with the generic GM model spectrum, 
it is possible that the  larger difference in energy levels
at eddy rim and center in the model would decrease over time similar to the observations, including appropriate forcing and other components
neglected so far in \eq{eq2.3.3}.
We also compare the spectra for single column and eddy simulations, where the additional horizontal shear in the eddy simulations leads to elevated energies for the center and rim.

We also observe prominent elevated IGW energies at the frequency $\omega=2.5 \times 10^{-3} \, \rm s^{-1}$ (or $log_{10}\omega f^{-1}\approx 1.9$) in the observations both at rim and center. 
This frequency is in fact the smallest value to which the Brunt-Väisälä frequency converges and therefore suggest that it is a distinct prevalent turning point. Although observations show a peak in energy at this frequency, our simulations do not model this peak, and so we conclude, that it may be caused by other wave processes.

Lastly, we compare the slopes of the observations (where the tidal signal has been filtered out) and model results fitted with a linear regression. The slopes fitted up to the frequency $\omega=2.5 \times 10^{-3} \, \rm s^{-1}$ show a very good agreement with a mean slope of $s=-1.2$ ($R^{2}=0.9$), where the single column simulations show the biggest difference due to the missing horizontal wave processes. Recently, \cite{dong_2023} also approximate a stationary energy spectrum that satisfies a power law with slope $s=-1$ from simulations based on the ray-tracing equations. 
The fits show steeper slopes for higher frequencies that range from $s=-2.1$ to $-2.4$.

\section{Summary and discussion}

A novel numerical approach based on the radiative transfer equation  is presented to study 
the propagation and refraction of IGWs in a meso-scale eddy in the ocean. 
In its full complexity the model includes six dimensions, three in physical and three in wavenumber space, 
in addition to  the time dimension. 
The  simulations presented here are the first ones, to our knowledge, to comprehensively 
describe the evolution of the full IGW spectrum inside a meso-scale coherent eddy, within the WKBJ-approximation,
 and detail the effects of wave  propagation, reflection, refraction, interaction 
with the mean flow, and breaking related to wave dissipation by vertical and horizontal refraction. 

Stratification and mean flow for the simulation are motivated from the observations of
the coherent cyclonic eddy structure $CE\_2019\_14N\_25W$ in the Canary Current System,
which we think is representative for a typical 
coherent eddy at mid-latitudes.
Mean flow and stratification are assumed to be stationary during a six day  simulation 
in which the wave field evolves and interacts with the background field, but no additional wave forcing is applied.
We use the GM model spectrum as  the initial condition. 
A reflection boundary condition at the  bottom and surface is used and open boundaries 
are implemented in the lateral direction, where energy freely flows out of the model domain, but no energy flows in.

In wavenumber space, we use also open boundaries at the  wavenumber thresholds
 $k_z^c=\pm 1 \, \rm m^{-1}$ vertically and  $k_h^c=\pm 0.5 \, \rm m^{-1}$ horizontally, and interpret the fluxes across 
 $k_z^c$ and $k_h^c$ towards smaller wavelength as wave breaking and dissipation by vertical and horizontal refraction, respectively.
 For given $\mathcal E$ from the GM model spectrum, $k_z^c$
is large enough for waves to be considered as unstable with  Ri$_{\mathrm{Bulk}}^{-1}\geq 1$ 
and thus are prone to breaking at $k_z^c$. The vertical wavenumber threshold $k_z^c$ is not uniquely given in literature, although a slightly larger vertical wavelength of $\lambda_c=10 \, \rm m$ (i.e. $k_z^c=0.63 \, \rm m^{-1}$)  has been reported previously (cf. \citet[]{gargett_1981, gregg_1987}). We test the impact of the choice $k_z^c=0.63 \, \rm m^{-1}$ on our results  using  single column simulations initialised with identical horizontal wavenumber space,  wave energy (the GM model),  mean flow corresponding to the northern eddy rim and stratification,
but different $k_z^c$.
Reducing the vertical wavenumber cutoff from $k_z^{c}=1 \, \rm m^{-1}$ to $k_z^{c}=0.63\, \rm m^{-1}$ leads to a $1.4$ times  larger wave energy loss after 6 days, by wave-mean flow interaction and dissipation. 
Furthermore, the resulting diffusivities are enhanced by an averaged value of $1.5$ times.
Note that only constructively superimposing waves are considered here, thus a condition 
Ri$_{\mathrm{Bulk}}^{-1}>4$ may  overestimate the breaking.
However, Ri$_{\mathrm{Bulk}}^{-1}$ depends on the total wave energy  that we use as  the initial condition, 
i.e. on the parameter $E_0$ of the GM model spectrum. 
With larger $E_0$, Ri$_{\mathrm{Bulk}}^{-1}$ will increase and $k_z^c$ will decrease. 
Note that three times larger values than used here for $E_0$  are found by \cite{PEO2017} 
 from ARGO float observations, i.e. the value for $E_0$ used here is most likely a low bias.

There is no known analogous treatise for $k_h^c$ as far as we know. 
Although we find that fluxes across $k_h^c$ are of two orders of magnitude 
smaller than fluxes across $k_z^c$, 
this may differ for cases with strong horizontally sheared mean flows. 
Thus, a proper assessment of horizontal  wavenumber thresholds may thus be required in such cases, 
but here we found the effects to be minor.

The  key results from this study are as follows:
\begin{itemize}
\item      Wave energy significantly decreases at the eddy rim, while the eddy center shows an increase in the wave energy.
We made an attempt to find such a difference in wave energy in the available observations,
which tend to show also larger wave energy in the eddy center, although with a much weaker
difference in the amplitudes.
 \item Waves gain energy from the mean flow outside of the eddy rim by lateral shear,
 while from eddy rim towards the center the effect vanishes.
 By vertical shear, the waves always lose energy to the mean flow, and this effect is largest
 at the eddy rim and towards the surface. 
 The net effect is to accelerate  the eddy inside 
 and to decelerate it outside its rim, with net
 deceleration overall.
 \item Lateral shear generates wave energy gain only due to a developing 
 horizontal anisotropy outside the eddy rim. This anisotropy is given by an 
 ellipse with maximal energy towards its center and its major axis oriented towards the eddy center, but slightly  tilted clockwise.
 The ellipse is generated in the first place by the effect of vertical refraction by vertical shear.
 The tilt is important for the effective wave energy gain and generated by horizontal refraction.
 \item Vertical shear extracts wave energy 
  by propagation into critical layers,
 without such dissipation wave-mean flow interaction by vertical shear has no net effect
 because of vertical reflection.
\item The effect of lateral wave propagation is as large as the mean-flow interaction 
and partly counterbalancing it,
but is more inhomogeneous in its pattern and more difficult to understand.
\item The effect of vertical propagation tends to be a redistribution of energy,
but at the eddy rim a loss of wave energy because of transports towards critical layers.
\item Wave breaking related to dissipation by vertical refraction due to vertical shear is two orders of magnitude larger
than the effect of horizontal shear (dissipation by horizontal refraction).

\item Wave breaking and dissipation by vertical refraction occurs predominantly at the eddy rims
close to the surface where vertical shear and changes in $N$ are largest,
and yields overall 2.3 \% energy loss over the integration period. 
The effect of $\p_z \v{U}_h$ dominates
but $\p_z N$ can also drive a minor energy loss
by dissipation by vertical refraction.

\item Related vertical diffusivities are also largest at the eddy rims
close to the surface and smallest at the eddy center and decay with depth,
with maximal values ranging from $\kappa = 10^{-5} \, \rm m^2s^{-1}$ at the start to
 $\kappa =  10^{-7} \, \rm m^2s^{-1}$ at the end of the simulation. These mixing rates appear low but note
 that they  depend on our choice of $E_0$ of the
 GM model spectrum we use as initial condition, which is likely to have a low bias.
\item The anticyclonic eddy behaves similar to the cyclonic one with respect to the results mentioned above, except
for the slight tilt of the horizontal anisotropy 
which is anticlockwise, opposite to the cyclonic eddy.
\end{itemize}

Dissipation measurements within the coherent eddy in the Canary Current system taken during the observational campaign will be used in the future 
to compare with the predicted enhanced mixing at the eddy rim, but are not yet available to us.
At the rim of a coherent meso-scale eddy in the Lofoten Basin,
\cite{fer2018dissipation} indeed find 
enhanced dissipation rates by turbulent mixing and relate that to internal wave breaking.
However, they also find enhanced dissipation at the eddy center, in contradiction to our model simulation. 
This difference might be related 
to different energy sources such as the instability of the background shear which \cite{fer2018dissipation} also find, or to the many limitations of our model approach.

The following limitations are inherent to our approach.
Wave propagation and refraction lead to numerical grid dispersion errors 
which are getting smaller using finer grid resolution,
but fine grids quickly become computationallvery expensive for our model. 
The simulations presented here are already at the upper end of available resources to us
such that using finer grids are not possible at the moment.
However, since the spectra and spatial distributions we simulate here remain continuous without large gradients on the small scales where the numerical errors are largest, we regard the effect of 
numerical grid dispersion errors as minor. 
The issue could be tested using some kind of self-refining 
or unstructured grids (triangular domain decomposition may have better properties than the rectangular grid used here).

Because of the assumed stationarity of the mean flow and stratification, 
the wave field might continuously build up energy at locations where the wave-mean flow interaction 
is strong despite the large energy sinks by wave-mean flow interaction or wave breaking. 
The missing implementation of the mean flow's reaction related to the wave energy changes
by wave-mean flow interaction as well as changes in stratification 
caused by induced mixing due to wave breaking, 
can lead to an unrealistic representation of the evolution of the wave field. 
However, without implementation of an interactive mean flow and stratification
these effects are difficult to assess and left for future studies. 
Further, the mean flow is not in balance with the background density field in our simulations, but we regard this effect as minor since the refraction by changing density should be dominated by the background vertical gradient $N^2(z)$.
Both issues could be assessed by comparing with a numerical simulation in physical space of waves interacting with a balanced eddy flow. There, however, other processes as wave-wave interaction and wave scattering at the mean flow and stratification are also present,
but which have been neglected in the present model simulations.

Recently, \cite{savva_kafiabad_vanneste_2021} derived (for $N=const$) a kinetic equation using the Wigner transform \citep{RYZHIK1996327}
similar to \eq{eq2.3.3},
which avoids the WKBJ assumption of slowly varying background conditions.
In contrast, 
this assumption  is necessary in our approach using the spectral energy balance \eq{eq2.3.3}, but note that
 we tested the WKBJ assumption in \fig{fig:wkb}
without finding significant violations.
The kinetic equation by  \cite{savva_kafiabad_vanneste_2021}
could in principle also be used instead 
of the spectral energy balance \eq{eq2.3.3}
(although a realistic $N(z)$ from observations
as used here could not be used).
In the kinetic equation by  
\cite{savva_kafiabad_vanneste_2021}
 the transport terms in wavenumber space and the energy transfer to or from the mean flow in  \eq{eq2.3.3} are replaced by a scattering term involving all other wavenumbers.
 Practically, 
 this scattering integral over all wavenumbers  would drastically
slow down a parallel integration since for each time step a sum over the (parallelized) wavenumber space
needs to be performed.
Conceptually, it is not obvious how to implement dissipation in the kinetic equation by \cite{savva_kafiabad_vanneste_2021} since the fluxes across wavenumber boundaries are missing there and are replaced by the scattering integral.
In contrast, \eq{eq2.3.3} is easily parallelized and 
dissipation is naturally implemented as described above.
A possible modification of IWEM to  the kinetic equation by \cite{savva_kafiabad_vanneste_2021}  is therefore postponed for now to a later study.

Wave forcing by tidal flow at the bottom or oscillatory Ekman pumping at the surface
appears simple  to implement to our model. 
Here, we ignore these source terms
and consider idealized spin-down simulations only, because of the other limitations
which we also currently not resolve. 
Another possible forcing which is more difficult 
to include is the secondary wave generation during the wave breaking process.
Currently, we assume that all the wave energy propagating to smaller wavelength
across the thresholds $k_z^c$ and $k_h^c$
is converted entirely to small-scale turbulence. However, a certain amount 
of the energy will certainly generate other waves at other wavelengths during
the breaking process, but this process is largely unknown and needs further attention.

Another limitation of our study is that as initial condition we use the generic GM model spectrum, but inside the eddy the spectrum is changing such that it is likely that  in a stationary state the wave field will differ from GM inside the eddy.
We found that even in single column simulations (not shown) with vanishing mean flow but depth-dependent stratification, 
the initial  GM model spectrum is not stationary but evolves in time due to vertical wave propagation and refraction.
This effect is related to the scaling of the GM model spectrum by $N$, i.e. by the term $E(z) = E_0 N(z)/N(0)$ in \eq{N123}.
The GM model spectrum only remains stationary for $N=const$ in our model.
Similar to \cite{Gregg_1989}, different power laws of $N$ for the formulation (larger and smaller then one)
have been tested for stationarity of the GM model spectrum, 
and it was found that other descriptions even increase the non-stationarity. 
Thus, the description of the GM model spectrum by the scaling $\mathcal{E}\sim N(z)/N_0$ 
appears to be a reasonable zero order approximation, but may not be the best representation \citep{Garrett1975S}. 
In fact, the observed GM model spectrum might be far from stationary such that
relatively large wave sources, transports and sinks are needed to explain the observations.
It is likely that for the 
GM model spectrum with its power law  of $-2$
the downscale energy transfers by non-linear wave-wave interactions also play 
an important role.
Such transports  are also not considered here and including a full evaluation 
of the scattering integral as in \citet{eden2019numerical}
 appears in principle possible but computationally  expensive. Some 
kind of approximation as in 
\citet{hasselmann1985computations} 
might help in this respect and is a subject of future work.

\acknowledgments
We thank Oliver B\"uhler and another anonymous reviewer for their comments and suggestions on this manuscript. 
This paper is a contribution to subprojects W6 and W2 of the Collaborative Research Centre 
TRR 181 ``Energy Transfers in Atmosphere and Ocean'' funded by the Deutsche Forschungsgemeinschaft 
(DFG, German Research Foundation) under project number 274762653. 
This paper is also associated with the research project REEBUS (Role of Eddies in the Carbon Pump 
of Eastern Boundary Upwelling Systems) funded by the Bundesministerium für Bildung 
und Forschung (BMBF, German Federal Ministry of Education and Research).
We thank M.~Dengler  and J.~Karstensen from GEOMAR, Kiel for providing 
the observational data and collaboration in REEBUS.

\datastatement
The novel numerical spectral Internal Wave Energy Model (IWEM) can be accessed via \url{https://github.com/ceden/IWEM.git}. 
The observational data used is taken from the data portal of the Helmholtz Centre for Ocean Research Kiel - GEOMAR, with restricted access.

\appendix[A] 
\appendixtitle{The Internal Wave Energy Model (IWEM)}
The Internal Wave Energy Model (IWEM) integrates the six-dimensional 
Radiative Transfer Equation \eq{eq2.3.3} in time, 
and simulates the interaction of the IGW field with the background medium. 
The model describes the propagation and refraction of IGWs in a variable background 
stratification and mean flow and the energy exchange of the wave field with the mean flow. 

Using a  staggered C-grid discretisation, wave energy $E$, stratification $N$ 
and mean flow velocities $U,V,W$ take values at the center of the grid boxes, while
zonal, meridional and vertical group velocities are located at the 
eastern, northern and upper side of the grid boxes. 
For the wave refraction in $k_x,k_y,k_z$-direction the discretisation is analogous. 
We use  an equidistant grid in the meridional direction, a surface refined 
non-equidistant grid in vertical direction, and also a non-equidistant grid in
wavenumber space, with finer resolution for small wavenumber components.
For parallelisation, each processor unit solves  \eq{eq2.3.3} on a subdomain of the model.

At the surface and bottom, 
waves are reflected back, so that upward propagating waves with $k_z<0$ propagate 
downward with $k_z>0$ after reflection and vice versa. 
All other boundaries are open, so that energy can freely flow out of the domain, but no energy can propagate into the domain.
For the computation of fluxes a necessary amount of numerical diffusion is introduced, 
so that energies remain positive definite. All fluxes are therefore computed with a second order advection scheme 
with superbee flux limiter  and forward time-stepping.
The numerical model code can be accessed at \url{https://github.com/ceden/IWEM.git}.

\appendix[B] \label{appendix:B2}
\appendixtitle{Observed velocities}

The observed zonal velocities in the meridional eddy cross section reach down to 1000 m depth. 
This is insufficient for a complete representation of the evolution of the IGW field, 
for which  we need 
 the mean flow over the whole depth. 
 Moreover, the observed velocities may 
 include  IGW signals which we need to remove.  
To  approximate the mean flow to the whole depth and to smooth the observations, 
a projection on the vertical modes is used. 
The orthogonal vertical eigenfunctions $\Phi_n$ result from the vertical eigenvalue problem
\begin{equation} \label{phi}
    \frac{\mathrm{d} }{\mathrm{d} z}f^{2}N^{-2}\frac{\mathrm{d} \Phi _{n}}{\mathrm{d} z}+f^{2}c_{n}^{-2}\Phi_{n}=0 \ \ \mbox{with} \ \ \frac{\mathrm{d} \Phi_n}{\mathrm{d} z} =0 \ \  at \ \ z=0, -h
\end{equation}
with approximate solution
\begin{equation}
\Phi_{n}\approx \sqrt{\frac{2N}{h\widetilde{N}}}cos \left ( \int_{-h}^{z} N/c_{n} \mathrm{d} {z}'\right ), \ \ c_{n}\approx h\widetilde{N}/(n\pi), \ \ \widetilde{N}=\frac{1}{h}\int_{-h}^{0}N\mathrm{d} z
\end{equation}
Velocity $U$ is then given by 
$U=\sum_{n=0}^{\infty }u_{n}\Phi_{n}(z)$ 
with $u_{n}=\int_{-h}^{0}U \Phi_{n}\mathrm{d} z$.
Since $U$ is not available for the whole depth down to 4000 m, the amplitudes $u_n$ for each mode are fitted for the available observations in the upper 1000 m of the ocean. 
The  eigenfunctions are calculated for $\overline{N}$ and the fitted profiles  $N_1(z)$ and $N_2(z)$. 
We carry out the fit with the Python function $scipy.optimize.curve\_fit$ first for the barotropic mode, 
subtract then the resulting velocities from the observed  ones 
and follow this procedure iteratively up to the 10$^{th}$ baroclinic mode, to obtain the mean flow.
This is done for each meridional position along the eddy cross section.
The barotropic and the first three baroclinic modes are the most powerful ones. 
The fit of these four modes results in a Pearson coefficient of $\rho=0.8$, 
hence the fit and the observed velocities correlate very well.  
The velocities resulting from the sum of the first 4 modes represent 
hence most of the mean flow and are used for the eddy simulations.

\appendix[C] \label{appendix:B3}
\appendixtitle{Idealized velocities}

Following, for example \cite{castelao_2011} and \cite{fischer_2021}, eddies captured in the tropical Atlantic can be well described by the following idealized structure:
 exponential decay outside of the eddy and a linear transition in-between eddy rims. 
 The eddy rim is located at a radius $r=\sqrt{x_{rim}^2+y_{rim}^2}=45 \, {\rm km}$ from the eddy center at $(x,y)_{center}=(120,120)\, {\rm km}$.
For the vertical direction, the first baroclinic mode $\Phi_1(z)$ from Appendix B is used. 
\begin{eqnarray}\label{velos}
U(x,y,z)&=-0.2\sqrt{L_x^{2}+L_y^{2}} \Phi_1(z) R(x,y) \sin(\tan^{-1} y/x) \\
V(x,y,z)&=0.2\sqrt{L_x^{2}+L_y^{2}} \Phi_1(z) R(x,y) \cos(\tan^{-1} y/x) \\
R(x,y)&=\left\{ \begin{array}{ll}
\sqrt{x^2+y^2}  
   & \mbox{for}~ 
   \sqrt{x^2+y^2} < r
   \\
     r  \exp\left( -\left( \sqrt{x^2+y^2} -r\right)/2r  \right)
    &  \mbox{for} ~
    \sqrt{x^2+y^2} > r 
\end{array} \right.
\end{eqnarray}
where $L_x$ and $L_y$ denote the horizontal extent of the domain.

\bibliographystyle{ametsocV6}
\bibliography{references}

\begin{thebibliography}{37}
\providecommand{\natexlab}[1]{#1}
\providecommand{\url}[1]{\texttt{#1}}
\renewcommand{\UrlFont}{\rmfamily}
\providecommand{\urlprefix}{URL }
\expandafter\ifx\csname urlstyle\endcsname\relax
  \providecommand{\doi}[1]{https://doi.org/\discretionary{}{}{}#1}\else
  \providecommand{\doi}{https://doi.org/\discretionary{}{}{}\begingroup
  \urlstyle{rm}\Url}\fi
\providecommand{\eprint}[2][]{\url{#2}}

\bibitem[{Browning et~al.(2021)Browning, Al-Hashem, Hopwood, Engel, Belkin,
  Wakefield, Fischer,, and Achterberg}]{fischer_2021}
Browning, T., A.~Al-Hashem, M.~Hopwood, A.~Engel, I.~Belkin, E.~Wakefield,
  T.~Fischer, and E.~Achterberg, 2021: {Iron regulation of North Atlantic eddy
  phytoplankton productivity}. \textit{Geophysical Research Letters},
  \textbf{48}, \doi{10.1029/2020GL091403}.

\bibitem[{B\"uhler and McIntyre(2005)B\"uhler, and McIntyre}]{buhler_2005}
B\"uhler, O., and M.~E. McIntyre, 2005: Wave capture and wave–vortex duality.
  \textit{Journal of Fluid Mechanics}, \textbf{534}, 67–95,
  \doi{10.1017/S0022112005004374}.

\bibitem[{Cairns and Williams(1976)Cairns, and Williams}]{gm76}
Cairns, J.~L., and G.~O. Williams, 1976: Internal wave observations from a
  midwater float, 2. \textit{Journal of Geophysical Research (1896-1977)},
  \textbf{81~(12)}, 1943--1950, \doi{https://doi.org/10.1029/JC081i012p01943}.

\bibitem[{Castelão and Johns(2011)Castelão, and Johns}]{castelao_2011}
Castelão, G.~P., and W.~E. Johns, 2011: {Sea surface structure of North Brazil
  Current rings derived from shipboard and moored acoustic Doppler current
  profiler observations}. \textit{Journal of Geophysical Research: Oceans},
  \textbf{116~(C1)}, \doi{10.1029/2010JC006575}.

\bibitem[{Chouksey et~al.(2018)Chouksey, Eden,, and
  Br{\"u}ggemann}]{chouksey_2018}
Chouksey, M., C.~Eden, and N.~Br{\"u}ggemann, 2018: {Internal gravity wave
  emission in different dynamical regimes}. \textit{Journal of Physical
  Oceanography}, \textbf{48}, \doi{10.1175/JPO-D-17-0158.1}.

\bibitem[{Chouksey et~al.(2022)Chouksey, Eden,, and Olbers}]{chouksey_2022}
Chouksey, M., C.~Eden, and D.~Olbers, 2022: Gravity wave generation in balanced
  sheared flow revisited. \textit{Journal of Physical Oceanography},
  \textbf{52~(7)}, 1351--1362.

\bibitem[{Dong et~al.(2023)Dong, Bühler,, and Smith}]{dong_2023}
Dong, W., O.~Bühler, and K.~S. Smith, 2023: Geostrophic eddies spread
  near-inertial wave energy to high frequencies. \textit{Journal of Physical
  Oceanography}, \textbf{53~(5)}, 1311 -- 1322,
  \doi{https://doi.org/10.1175/JPO-D-22-0153.1},
  \urlprefix\url{https://journals.ametsoc.org/view/journals/phoc/53/5/JPO-D-22-0153.1.xml}.

\bibitem[{Duda and Cox(1989)Duda, and Cox}]{duda_cox_1989}
Duda, T.~F., and C.~S. Cox, 1989: {Vertical wave number spectra of velocity and
  shear at small internal wave scales}. \textit{Journal of Geophysical
  Research: Oceans}, \textbf{94~(C1)}, 939--950, \doi{10.1029/JC094iC01p00939}.

\bibitem[{Eden et~al.(2019)Eden, Pollmann,, and Olbers}]{eden2019numerical}
Eden, C., F.~Pollmann, and D.~Olbers, 2019: Numerical evaluation of energy
  transfers in internal gravity wave spectra of the ocean. \textit{Journal of
  Physical Oceanography}, \textbf{49~(3)}, 737--749.

\bibitem[{Fer et~al.(2018)Fer, Bosse, Ferron,, and
  Bouruet-Aubertot}]{fer2018dissipation}
Fer, I., A.~Bosse, B.~Ferron, and P.~Bouruet-Aubertot, 2018: The dissipation of
  kinetic energy in the lofoten basin eddy. \textit{Journal of Physical
  Oceanography}, \textbf{48~(6)}, 1299--1316.

\bibitem[{Garanaik and Venayagamoorthy(2019)Garanaik, and
  Venayagamoorthy}]{garanaik_venayagamoorthy_2019}
Garanaik, A., and S.~K. Venayagamoorthy, 2019: {On the inference of the state
  of turbulence and mixing efficiency in stably stratified flows}.
  \textit{Journal of Fluid Mechanics}, \textbf{867}, 323–333,
  \doi{10.1017/jfm.2019.142}.

\bibitem[{Gargett et~al.(1981)Gargett, Hendricks, Sanford, Osborn,, and
  Williams}]{gargett_1981}
Gargett, A.~E., P.~J. Hendricks, T.~B. Sanford, T.~R. Osborn, and A.~J.
  Williams, 1981: {A composite spectrum of vertical shear in the upper ocean}.
  \textit{Journal of Physical Oceanography}, \textbf{11~(9)}, 1258--1271,
  \doi{10.1175/1520-0485(1981)011<1258:ACSOVS>2.0.CO;2}.

\bibitem[{Garrett and Munk(1979)Garrett, and Munk}]{Garrett_Munk_1979}
Garrett, C., and W.~Munk, 1979: {Internal waves in the ocean}. \textit{Annual
  Review of Fluid Mechanics}, \textbf{11~(1)}, 339--369,
  \doi{10.1146/annurev.fl.11.010179.002011}.

\bibitem[{Garrett and Munk(1975)Garrett, and Munk}]{Garrett1975S}
Garrett, C. J.~R., and W.~H. Munk, 1975: {Space-time scales of internal waves'
  a progress report}. \textit{Journal of Geophysical Research}, \textbf{80},
  291--297.

\bibitem[{Gregg(1987)}]{gregg_1987}
Gregg, M.~C., 1987: {Diapycnal mixing in the thermocline: A review}.
  \textit{Journal of Geophysical Research: Oceans}, \textbf{92~(C5)},
  5249--5286, \doi{10.1029/JC092iC05p05249}.

\bibitem[{Gregg(1989)}]{Gregg_1989}
Gregg, M.~C., 1989: {Scaling turbulent dissipation in the thermocline}.
  \textit{Journal of Geophysical Research}, \textbf{94}, 9686--9698.

\bibitem[{Hasselmann et~al.(1985)Hasselmann, Hasselmann, Allender,, and
  Barnett}]{hasselmann1985computations}
Hasselmann, S., K.~Hasselmann, J.~Allender, and T.~Barnett, 1985: {Computations
  and parameterizations of the nonlinear energy transfer in a gravity-wave
  specturm. Part II: Parameterizations of the nonlinear energy transfer for
  application in wave models}. \textit{Journal of Physical Oceanography},
  \textbf{15~(11)}, 1378--1391.

\bibitem[{Kafiabad et~al.(2019)Kafiabad, Savva,, and
  Vanneste}]{kafiabad_savva_vanneste_2019}
Kafiabad, H.~A., M.~A.~C. Savva, and J.~Vanneste, 2019: Diffusion of
  inertia-gravity waves by geostrophic turbulence. \textit{Journal of Fluid
  Mechanics}, \textbf{869}, R7, \doi{10.1017/jfm.2019.300}.

\bibitem[{Kunze and Smith(2004)Kunze, and Smith}]{KunzeSmith2004}
Kunze, E., and S.~Smith, 2004: {The role of small-scale topography in turbulent
  mixing of the global ocean}. \textit{Oceanography}, \textbf{17}, 55--64,
  \doi{10.5670/oceanog.2004.67}.

\bibitem[{M{\"u}ller and Briscoe(2000)M{\"u}ller, and
  Briscoe}]{mullerbriscoe_2000}
M{\"u}ller, P., and M.~Briscoe, 2000: {Diapycnal mixing and internal waves}.
  \textit{Oceanography}, \textbf{13}, \doi{10.5670/oceanog.2000.40}.

\bibitem[{Munk(1981)}]{Munk19819IW}
Munk, W.~H., 1981: {Internal waves and small-scale processes}. 264--269.

\bibitem[{Natarov and M{\"u}ller(2005)Natarov, and
  M{\"u}ller}]{muller_natarov_2005}
Natarov, A., and P.~M{\"u}ller, 2005: {A dissipation function for internal wave
  radiative balance equation}. \textit{Journal of Atmospheric and Oceanic
  Technology}, \textbf{22}, \doi{10.1175/JTECH1788.1}.

\bibitem[{Nikurashin and Ferrari(2011)Nikurashin, and Ferrari}]{ferrari_2011}
Nikurashin, M., and R.~Ferrari, 2011: {Global energy conversion rate from
  geostrophic flows into internal lee waves in the deep ocean}.
  \textit{Geophysical Research Letters}, \textbf{38},
  \doi{10.1029/2011GL046576}.

\bibitem[{Olbers and Eden(2013)Olbers, and Eden}]{olbers2013}
Olbers, D., and C.~Eden, 2013: {A global model for the diapycnal diffusivity
  induced by internal gravity waves}. \textit{Journal of Physical
  Oceanography}, \textbf{43}, \doi{10.1175/JPO-D-12-0207.1}.

\bibitem[{Olbers et~al.(2019)Olbers, Eden, Becker, Pollmann,, and
  Jungclaus}]{Olbers2019}
Olbers, D., C.~Eden, E.~Becker, F.~Pollmann, and J.~Jungclaus, 2019:
  \textit{{The IDEMIX Model: Parameterization of internal gravity waves for
  circulation models of ocean and atmosphere}}, 87--125. Springer International
  Publishing, Cham, \doi{10.1007/978-3-030-05704-6_3}.

\bibitem[{Olbers et~al.(2012)Olbers, Willebrand,, and Eden}]{olbers_eden_2012}
Olbers, D., J.~Willebrand, and C.~Eden, 2012: \textit{{Ocean dynamics}},
  211--286. \doi{10.1007/978-3-642-23450-7_8}.

\bibitem[{Osborn(1980)}]{Osborn_1980}
Osborn, T.~R., 1980: {Estimates of the local rate of vertical diffusion from
  dissipation measurements}. \textit{Journal of Physical Oceanography},
  \textbf{10~(1)}, 83 -- 89,
  \doi{10.1175/1520-0485(1980)010<0083:EOTLRO>2.0.CO;2}.

\bibitem[{Osborn and Cox(1972)Osborn, and Cox}]{Osborn_Cox_1972}
Osborn, T.~R., and C.~S. Cox, 1972: {Oceanic fine structure}.
  \textit{Geophysical Fluid Dynamics}, \textbf{3~(4)}, 321--345,
  \doi{10.1080/03091927208236085}.

\bibitem[{Pollmann et~al.(2017)Pollmann, Eden,, and Olbers}]{PEO2017}
Pollmann, F., C.~Eden, and D.~Olbers, 2017: Evaluating the global internal wave
  model idemix using finestructure methods. \textit{Journal of Physical
  Oceanography}, \textbf{47~(9)}, 2267 -- 2289, \doi{10.1175/JPO-D-16-0204.1}.

\bibitem[{Polzin(2004)}]{polzin2004}
Polzin, K., 2004: {Idealized solutions for the energy balance of the finescale
  internal wave field}. \textit{Journal of Physical Oceanography}, \textbf{34},
  \doi{10.1175/1520-0485(2004)034<0231:ISFTEB>2.0.CO;2}.

\bibitem[{Polzin(2009)}]{polzin2009}
Polzin, K., 2009: {An abyssal recipe}. \textit{Ocean Modelling}, \textbf{30},
  298--309, \doi{10.1016/j.ocemod.2009.07.006}.

\bibitem[{{REEBUS: Role of eddies in the carbon pump of the eastern boundary
  upwelling systems}(2019)}]{reebus}
{REEBUS: Role of eddies in the carbon pump of the eastern boundary upwelling
  systems}, 2019: {The importance of climate change in coastal upwelling
  areas}. GEOMAR, UHH.

\bibitem[{Ryzhik et~al.(1996)Ryzhik, Papanicolaou,, and Keller}]{RYZHIK1996327}
Ryzhik, L., G.~Papanicolaou, and J.~B. Keller, 1996: Transport equations for
  elastic and other waves in random media. \textit{Wave Motion},
  \textbf{24~(4)}, 327--370,
  \doi{https://doi.org/10.1016/S0165-2125(96)00021-2},
  \urlprefix\url{https://www.sciencedirect.com/science/article/pii/S0165212596000212}.

\bibitem[{Savva et~al.(2021)Savva, Kafiabad,, and
  Vanneste}]{savva_kafiabad_vanneste_2021}
Savva, M., H.~Kafiabad, and J.~Vanneste, 2021: Inertia-gravity-wave scattering
  by three-dimensional geostrophic turbulence. \textit{Journal of Fluid
  Mechanics}, \textbf{916}, A6, \doi{10.1017/jfm.2021.205}.

\bibitem[{Sherman(1989)}]{shermanphd_1989}
Sherman, J.~T., 1989: {Observations of fine-scale vertical shear and strain in
  the upper ocean}. Ph.D. thesis, University of California, San Diego.

\bibitem[{Sherman and Pinkel(1991)Sherman, and Pinkel}]{Sherman_Pinkel_1991}
Sherman, J.~T., and R.~Pinkel, 1991: {Estimates of the vertical
  wavenumber--frequency spectra of vertical shear and strain}. \textit{Journal
  of Physical Oceanography}, \textbf{21~(2)}, 292 -- 303,
  \doi{10.1175/1520-0485(1991)021<0292:EOTVWS>2.0.CO;2}.

\bibitem[{Wunsch and Ferrari(2003)Wunsch, and Ferrari}]{wunschferrari2003}
Wunsch, C., and R.~Ferrari, 2003: {Vertical mixing, energy, and the general
  circulation of the oceans}. \textit{Annu. Rev. Fluid Mech}, \textbf{18},
  281--314, \doi{10.1146/annurev.fluid.36.050802.122121}.

\end{thebibliography}

\end{document}